\documentclass[12pt]{iopart}   

\usepackage{iopams}
\usepackage[english]{babel}
\usepackage{graphicx}
\usepackage{dcolumn}
\usepackage{bm}
\usepackage{verbatim}
\usepackage{amsfonts}
\usepackage{amssymb}
\usepackage{color}
\usepackage{enumerate}
\newcommand{\be}[1]{\begin{equation}\label{#1}}
\newcommand{\ee}{\end{equation}}
\newcommand{\bea}[1]{\begin{eqnarray}\label{#1}}
\newcommand{\eea}{\end{eqnarray}}
\newcommand{\no}{\nonumber \\}
\newcommand{\Fig}[1]{Fig.(\ref{#1})}
\newcommand{\Eq}[1]{Eq.(\ref{#1})}

\newcommand{\Sec}[1]{Section~\ref{#1}}
\newcommand{\Tbl}[1]{Table~\ref{#1}}
\newcommand{\om}{\omega}

\newcommand{\half}{\frac{1}{2}}
\newcommand{\thalf}{\tiny{\frac{1}{2}}}

\def\myoverDefn#1#2{\hbox{\space \raise1mm\hbox{ $\scriptstyle{#1} \atop \scriptstyle{#2}$} }}

\def\np{{n_p}}
\def\np0{{n_{p0}}}

\def\asdag{a_s^\dagger}

\def\nbarp{\bar{n}_p}
\def\nbars{\bar{n}_s}
\def\nbari{\bar{n}_i}
\usepackage{color}

\def\trm#1{\textrm{#1}}
\def\tit#1{\textit{#1}}
\def\tbf#1{\textbf{#1}}

\def\tfrac#1#2{{\textstyle \frac{#1}{#2}}}
\def\thalf{{\textstyle \frac{1}{2}}}
\def\tfourth{{\textstyle \frac{1}{4}}}

\newcommand{\ket}[1]{|#1\rangle}
\newcommand{\bra}[1]{\langle #1|}
\newcommand{\pd}{\partial}

\def\SBekHawk{S_{\trm{\tiny Bekenstein-Hawking}}}
\def\quarter{\frac{1}{4}}
\def\Tr{\trm{Tr}}
\newcommand{\appropto}{\mathrel{\vcenter{
  \offinterlineskip\halign{\hfil$##$\cr
    \propto\cr\noalign{\kern2pt}\sim\cr\noalign{\kern-2pt}}}}}
\newcommand{\defn}{\mathrel{\vcenter{
  \offinterlineskip\halign{\hfil$##$\cr
    {\scriptsize \trm{def}}\cr\noalign{\kern2pt}=\cr\noalign{\kern-2pt}}}}}
\newcommand{\mylimit}[2]{\mathrel{\vcenter{
  \offinterlineskip\halign{\hfil$##$\cr
    {\scriptstyle {#1}}\cr\noalign{\kern2pt}{#2\;\;}\cr\noalign{\kern-2pt}}}}}    

\begin{document}
\title[Black Hole Waterfall: a unitary phenomenological model for black hole evaporation ]{Black Hole Waterfall: a unitary phenomenological model for black hole evaporation with Page curve}
\author{P.M. Alsing }
\address{Florida Atlantic University, 777 Glades Rd, Boca Raton, FL, 33431}
\ead{palsing@fau.edu,alsingpm@gmail.com}
\date{\today}

\begin{abstract}
We present a unitary phenomenological model for black hole evaporation based on the analogy of the  laboratory process of spontaneous parametric down conversion (SPDC) \cite{Alsing:2015,Alsing:2016} when the black hole (pump) is allowed to deplete to zero mass.  The model incorporates an additional new feature that allows for the interior Hawking partner-particles (idlers) behind the horizon to further generate new Hawking particle pairs of lower energy, one of which remains behind the horizon, and the other  that adds to the externally emitted Hawking radiation (signals) outside the horizon. This model produces a Page curve for the evolution of the reduced density matrices for the evaporating black hole internal degrees of freedom entangled with the generated Hawking radiation pairs entangled across the horizon.
The Page curve yields an entropy that rises at early times during the evaporation process as Hawking pairs are generated, reaches a peak midway through the evolution, and then decays to zero upon complete evaporation of the black hole.
 The entire system remains in a pure state at all times undergoing unitary (squeezed state) evolution, with the initial state of the black hole modeled  as a bosonic Fock state of large, but finite number $\np0$ of particles. For the final state of the system, the black hole reaches the vacuum state of zero mass, while the external Hawking radiation carries away the total energy of the initial black hole. Inside the horizon there remains  $\np0$ Hawking partner-particles of vanishingly small total energy, reminiscent of the "soft-hair" (zero energy) qubit model of Hotta, Nambu and Yamaguchi \cite{Hotta_Nambu_Yamaguchi:2018}, but now from a Hamiltonian for squeezed state generation perspective. 
 The model presented here can be readily extended to encompass arbitrary  initial pure states for the black hole, and in falling matter.
%
\end{abstract}

\section{Introduction}\label{Intro}
In the last two decades great progress has been made in a deeper understanding of the origin of the entropy of the Hawking radiation due to the evaporation of a black hole (BH), the resulting black hole information problem (paradox), and its intimate connection to quantum entanglement. 
Before we address the unitary phenomenological models presented in this work, let us first remind ourselves of the central thorny issues raised BH evaporation, and discuss the recent developments that have arisen to address them. 
In introductory sections below we review the Black Hole Information Problem (Paradox), the Page Curve, and the modern view of Black Hole Evaporation that has emerged over the last two decades.
We end by stating the goals and objectives that should be achieved by any model of BH evaporation.
In \Sec{sec:ModelReview}
 review the unitary phenomenological models and results presented in \cite{Alsing:2015,Alsing:2016} for BH evaporation evolving from an explicit pure state, and discuss 
 their achievements and deficiencies in creating a Page Curve (i.e. BH and Hawking radiation curves that initially increase in early times, reach zero slope some midway point (the ``Page time"), and then proceed to decrease to zero, with the BH fully evaporated). 
 We then introduce the extension of those models in \Sec{sec:BHWModel}, which we term the BH Waterfall model, in which the internal Hawking partner particles, generated by the pump/BH, can themselves act as further pump sources for subsequent spontaneous parametric down conversion (SPDC) two-mode squeezed state vacuum generation \cite{Agarwal:2013,Gerry_Knight:2023}. We show how this model can achieve the desired stated goals of a BH evaporation process.

\subsection{The Black Hole Information Problem in a Nutshell}\label{sec:BHIP}
Briefly, if the BH begins in a pure state, and the emitted Hawking radiation is thermal \cite{Hawking:1975}, then these conditions are at odds with the decreasing Bekenstein-Hawking thermodynamic entropy
given by $\SBekHawk = \quarter \frac{A_{BH}}{L^2_{Pl}} =  4 \pi (\frac{M}{M_{Pl}})^2$, where 
$A_{BH} = 4 \pi r^2_s$ is the area of the horizon of a Schwarzschild BH of radius $r_s = 2 G M/c^2$ of mass $M$, and  $L^2_{Pl} = \hbar G/c^3$ and $M^2_{Pl} = \hbar c/G$ are the squares of the Planck length and Planck mass, respectively. Essentially, the issue stems from the following considerations. The thermal nature of the Hawking radiation arises from the highly entangled nature of the vacuum, which in the case of the maximally extended Schwarzschild spacetime is envisioned to be the thermofield double (TFD) state, 
$\ket{TFD} = Z^{-1} \sum_{n=0}^\infty e^{-\beta\,E_n}\,\ket{n}_L\otimes\ket{n}_R$ \cite{Almheiri_Hartman:2020,Almheiri_Hartman:2021}, where 
$Z = \sum_{n=0}^\infty e^{-\beta\,E_n}$  is the partition function, and $L$ and $R$ are the causally disconnected left and right Rindler wedges of the associated Penrose diagram. In the external region of the BH in ``our universe," (the R-wedge) the reduced quantum state of the of the Hawking radiation $\rho_{R}$ is given by tracing out over the causally disconnected region $L$ wedge, since no signal from the latter can enter the $R$ wedge. This yields 
$\rho_{R} = Z^{-1} \sum_{n=0}^\infty e^{-\beta\,E_n}\,\ket{n}_R\bra{n}$ which is a thermal state with Boltzmann probabilities $p_n = e^{-\beta\,E_n}/Z$, with accompanying von Neumann \tit{fine-grained}  entropy 
$S = -\Tr[\rho_{R}\log \rho_{R}] =  -\Tr[\sum_n p_n\log p_n]$. Finally, since the state is pure, the reduced density matrices  of the bipartite $BH\cup R$  (black hole/Hawking radiation) system are equal (having a common set of nonzero eigenvalues \cite{NC:2000,Wilde:2017}), $S(\rho_{BH}) = S(\rho_{R})$, where $S(\rho_{BH}) = \Tr_{R}[\rho_{BH,R}]$ and 
$S(\rho_{R}) = \Tr_{BH}[\rho_{BH,R}]$.

However, the thermodynamic Bekenstein-Hawking  entropy $\SBekHawk$ is a \tit{coarse-grained} entropy.
In general, this means \cite{Almheiri_Hartman:2021} that given a subset  $\mathcal{O}_i$ (a coarse graining) of all possible observables $\mathcal{O}$ of a system, we consider all  density matrices $\tilde{\rho}$ which give the same results for as our system $\rho$ for the observables we are measuring, 
$\Tr[\tilde{\rho}\,\mathcal{O}_i] = \Tr[\rho\,\mathcal{O}_i]$. One then computes the fine-grained von Neumann entropy $S(\tilde{\rho})$ and finally, maximizes over all possible choices of $\tilde{\rho}$. 
Thus, by definition, any fine-grained entropy is necessarily less than or equal to its coarse-grained upper bound. For the case of BH evaporation this implies that $S(\rho_R)\le\SBekHawk$. But this posses a problem since the area, and hence coarse-grained entropy of the BH, is shrinking during the evaporation process, while the fine-grained entropy of the Hawking radiation is increasing. 
After the point in time at which $S(\rho_{R})=\SBekHawk$ we have an inconsistency, since then
we'd have  $S(\rho_{R})>\SBekHawk$.
Further, this problem occurs much earlier than the time at which the BH completely evaporates, i.e. at a time when  the mass of the BH is roughly half its initial value, and the effects of gravity at the horizon are not necessarily strong.  Thus, a semiclassical calculation, such as performed by Hawking's using quantum field theory on a fixed classical general relativistic  spacetime, should be appropriate. Additionally, at the endpoint of BH evaporation process when only the Hawking radiation remains, the initial pure state of the system has evolved into a mixed state, which would violate the unitarity of quantum mechanics. 

\subsection{The Page Curve}\label{sec:ThePageCurve}
\begin{figure}[h]
\begin{center}
\includegraphics[width=4in,height=2.5in]{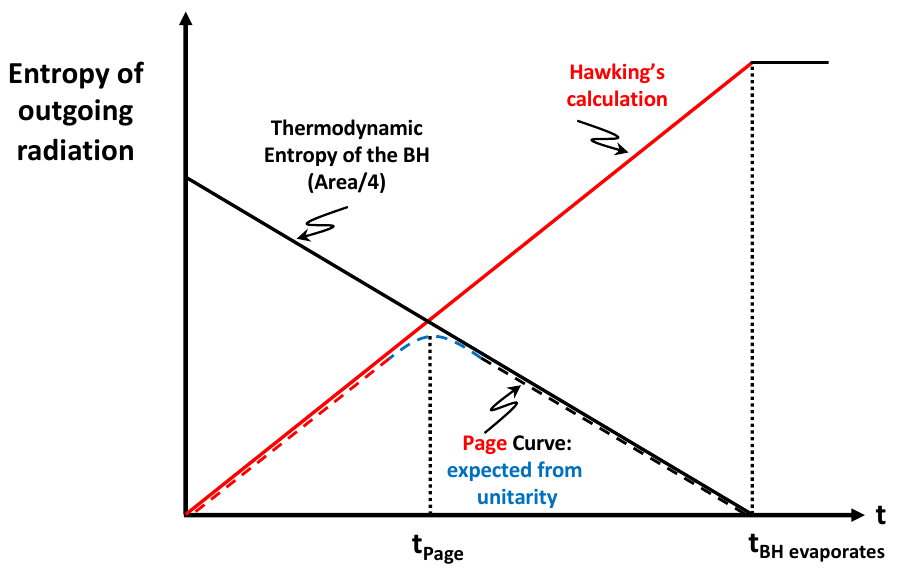} 
\end{center}
\caption{Page Curve.  
If unitarity is maintained, and hence information is not lost during the BH evaporation process, the evolution of the entropy of the black hole and the external Hawking radiation should follow the curved lower red-black curve. 
}\label{fig:PageCurve}
\end{figure}

If unitarity is to be maintained, and hence information is not lost,  and $S(\rho_{R})\le\SBekHawk$ at all times then something must happen at the crossover point when $S(\rho_{R})=\SBekHawk$. 
This is essentially Don Page's seminal argument \cite{Page:1993a,Page:1993b} leading to the celebrate Page curve depicted as the (lower) red-black dashed line
in \Fig{fig:PageCurve}. 
The crossover point is called the Page time $t_{Page}$. 
(Note: The rising red and falling black solid straight lines of entropy in \Fig{fig:PageCurve} computed by Hawking's calculation, and that given by the Bekenstein-Hawking theromodynamic entropy, respectively are merely representative). 
Before the Page time, as entangled interior/exterior Hawking pairs are just starting to be generated, the  entropy of the outgoing radiation is dominated by the fine-grained entropy of Hawking's calculation (rising red solid curve). After the Page time, when a substantial number of the Hawking pairs have been emitted and the BH has shrunk to roughly half its mass, the entropy of the Hawking radiation should follow the coarse-grained entropy  of the ever-decreasing area (divided by 4) of the BH (falling black solid curve). 
If unitarity is maintained, the evolution of the entropy of the 
Hawking radiation should follow the (lower) rising curve 
(red-dashed)  for  $t\le t_{Page}$, and then falling (black-dashed) curve for  $t\ge t_{Page}$), with BH evaporation as the end point of the evolution at some $t=t_{BH\,evaps}$. 

The  fate of the interior Hawking particles during the BH evaporation process,  
presents an additional issue as well.
If the BH begins in a pure state, then its entropy is clearly zero. If the BH entropy is also zero at 
$t=t_{BH\,evaps}$ then the BH  must return to a pure state. However, of the  Hawking pairs generated during the BH evaporation process, what precisely is the fate of the interior Hawking-partner particles? 
Because the Hawking radiation carries away energy, and the area of the BH $A_{BH}\sim M^2$ shrinks, the interior Hawking particles in effect carry negative energy which can reduce the mass of the BH inside the horizon. From a classical general relativistic energy conservation perspective \cite{Hartle:2003} one can argue as follows. If a Hawking pair of particles is created out of the vacuum on opposite sides of the horizon (for a stationary BH)  with four-momenta $\mathbf{p}$ and $\bar{\mathbf{p}}$ (exterior, interior, respectively) 
then the total energy 
$\mathbf{p}\cdot\boldsymbol{\xi} + \bar{\mathbf{p}}\cdot\boldsymbol{\xi}=0$, where 
$\boldsymbol{\xi} = -(1-2\,M/r)$ is the Killing vector for the Schwarzschild metric. If the exterior particle with energy  $\mathbf{p}\cdot\boldsymbol{\xi}> 0$ escapes to infinity, the interior particle must be considered to have negative ``energy"  $\bar{\mathbf{p}}\cdot\boldsymbol{\xi}<0$. This situation \tit{can} possibly occur because the Killing vector $\boldsymbol{\xi}$ changes sign inside the horizon $r<2M$ and hence becomes spacelike. (In this case, where $r$ is now the timelike coordinate  $\bar{\mathbf{p}}\cdot\boldsymbol{\xi}$ is really a momentum, which can be either positive or negative). The mass of the BH is reduced by the energy lost to the escaping exterior particle, which is the value  $-\bar{\mathbf{p}}\cdot\boldsymbol{\xi}>0$ of the interior particle.  However, from a quantum perspective this presents an additional problem.
If the interior Hawking particles are annihilated with an equivalent mass/energy 
$\Delta M=-\bar{\mathbf{p}}\cdot\boldsymbol{\xi}$  portion of the BH, then it would seem that the entanglement with its exterior Hawking partner is also destroyed.
Thus, if the interior Hawking particles 
 disappear along with the BH at $t=t_{BH\,evaps}$,  then what is the (external) emitted Hawking radiation entangled with (so that the trace over the interior produces a thermal state for the Hawking radiation)?
 
 \subsection{A brief review of the modern view of Black Hole evaporation}\label{sec:ModernView:BHEvap}
In the last two decades, intense scrutiny has been brought to bear on the above issues in an attempt to not only resolve the BH Information Problem (Paradox), but also to appreciate and understand what was missing from the original Hawking calculation. (For an  informative recent (2021) review see Almeheiri, Hartman  \tit{et al.} \cite{Almheiri_Hartman:2021} and references therein, and the (2016) Jerusalem Lectures by Harlow \cite{Harlow:2016}. For a  popular pedagogical, though insightful (2022) review see Cox and Forshaw \cite{Cox_Forshaw:2022} and references therein). The current belief of a majority of researchers  over the last forty years is that a BH can be regarded as an ordinary system obeying the laws of thermodynamics. More precisely, a BH can be described as an object with a finite, but large number of degrees of freedom that obey the ordinary laws of physics, including thermodynamics. As put by Almeheiri, Hartman  \tit{et al.} \cite{Almheiri_Hartman:2021} this ``Central Dogma," can be stated as ``As seen from the outside, a BH can be described in terms of a quantum system with Area/(4 G) degrees of freedom, which evolved unitarily under time evolution." (In unit where $\hbar = c =1$, which we use from now on, 
$G = L_{Pl}^2$, and $1/G = M_{Pl}^2$).

\begin{figure}[h]
\begin{center}
\begin{tabular}{ccc}
\hspace{-0.7in}
\includegraphics[width=3.95in,height=3.25in]{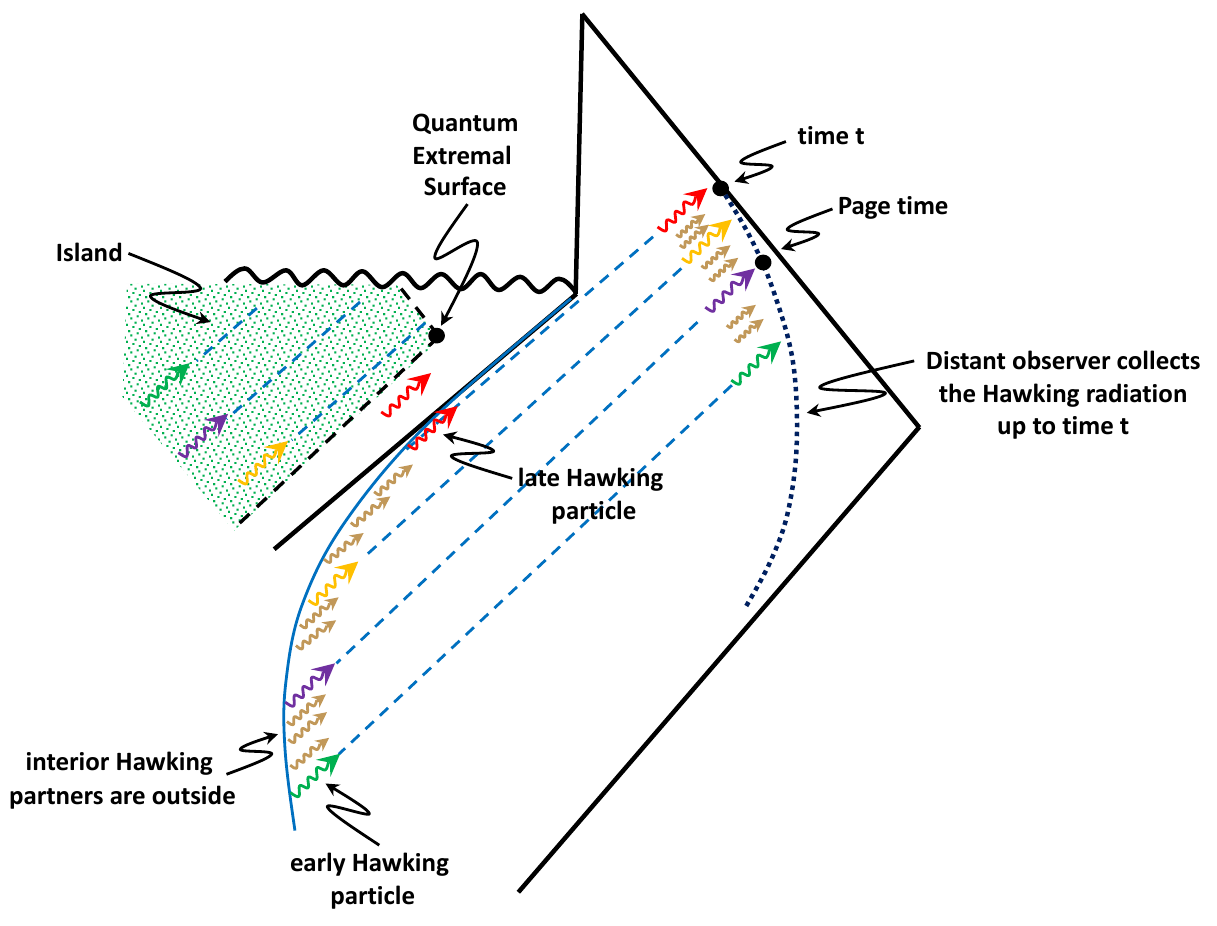} &
{\hspace{1em}} &
\hspace{-0.5in}
\includegraphics[width=3.75in,height=3.5in]{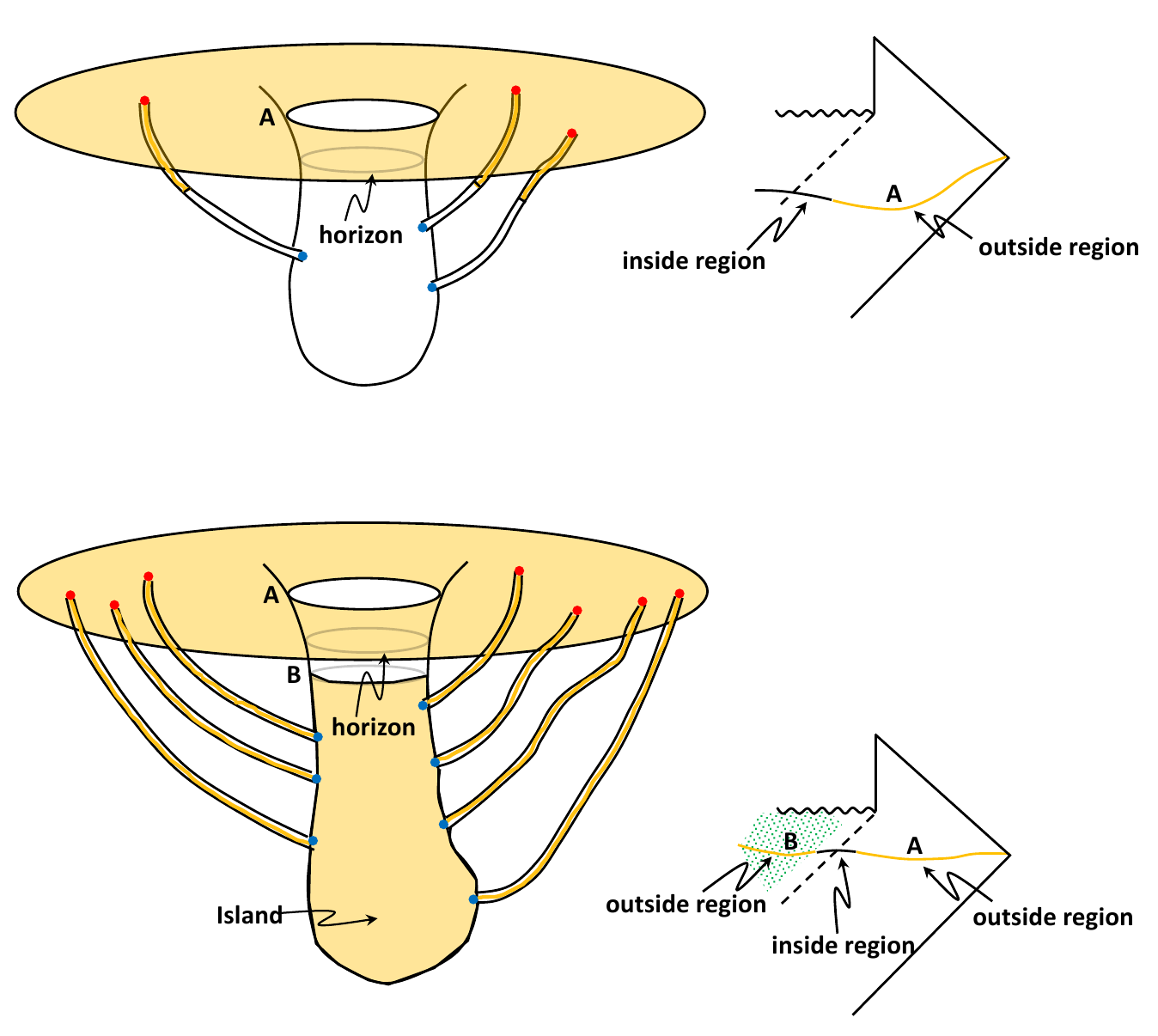}
\end{tabular}
\end{center}
\caption{(left) Penrose diagram for an evaporating BH, showing entangled particle pairs (matching colored arrows), the \tit{island} (shaded green area inside the horizon) and the quantum extremal surface (QES, black dot).
(right) Cartoon illustrating the concept that interior Hawking partner particles (blue dots) in the island behind the horizon are ``transferred" to the exterior Hawking radiation (red dots)  by a growing network of microscopic wormholes (tubular structures). (See main text, and \cite{Cox_Forshaw:2022, Penington:2020} for further details). 
}\label{fig:Cox:14.5:14.6}
\end{figure}

Evidence for such a belief has come from computation of the BH entropy for special extremal BHs in supersymmetric string theories, demonstrating that  an explicit count of microstates equals $\SBekHawk$.
These computations match not only the area formula, but also its corrections (see references in \cite{Almheiri_Hartman:2021}). Further evidence, which spurred much research continuing to this day, 
 was provided by  Maldacena's seminal work on the AdS/CFT correspondence.  
 The AdS/CFT correspondence conjectures a relation between the physics of Anti-deSitter (AdS) spacetime
and a dual conformal field theory (CFT) living on its boundary. In this case, the black hole and its whole exterior can be represented in terms of degrees of freedom living on the boundary, a concept known as Holography, introduced in 1995 by Susskind \cite{Susskind:1995}. 
(For more details, and applications beyond just gravity, see \cite{Natsuume:2012,Nastase:2015,Nastase:2017, Erdmenger:2017} and references therein). 
These gave rise to the concept of Holographic Entanglement  in the mid-2000s due to the development of the Ryu-Takayanagi (RT, 2006) conjecture/formula \cite{Ryu_Takayanagi:2006} and its subsequent covariant generalization by Hubney, Rangamani Takayanagi (HRT, 2007) \cite{HRT:2007}. The RT/HRT formula generalizes the Bekenstein-Hawking entropy formula $\SBekHawk = \quarter\,Area_{BH}$ to the area of a quantum extremal surface (QES), which typically lies just within the BH horizon, and whose contribution to the entanglement entropy replaces that of the area of the BH horizon. The entanglement entropy 
$S_R$ of the Hawking radiation is then given by the extremization of the \tit{generalized entropy} give by
\be{Cox:p249}
S_R = \frac{Area(QES)}{4} + S_{SC},
\ee
where $S_{SC}$ is the entanglement entropy of the Hawking radiation as computed by Hawking in the semiclassical (SC subscript) calculation, but now with a crucial difference. The calculation now mandates that we should also include in $S_{SC}$ some of the interior Hawking-particle partners \tit{inside the horizon} in what has come to be known as the \tit{island} (see the 2020 concurrently published papers of  Penington  \cite{Penington:2020}, and Almheiri, Engelhardt, Marolf and Maxfield \cite{Almheiri:2020} for more details, as well as the descriptive reviews by \cite{Almheiri_Hartman:2021,Cox_Forshaw:2022}).
 This is illustrated in \Fig{fig:Cox:14.5:14.6}(left), where the rightmost dashed blue line indicates the worldline of a 
distant observer collecting the exterior Hawking radiation at different times $t$. The red arrow outside the horizon depicts late time emitted exterior Hawking particles (radiation) which are entangled with their interior partner particles (also red) just inside the horizon. Early time emitted interior Hawking partner particles (various other colored arrows) that exist within the island inside the horizon are now considered \tit{as part} of the external Hawking radiation (matching colored arrows) in the computation of $S_{SC}$. 
This is accomplished by postulating the existence of microscopic wormholes which connect interior Hawking partner particles with the external Hawking radiation.
The function of the microscopic wormholes is to ``unite"  interior Hawking particles with the exterior partner particles in the external Hawking radiation, which then acts to purify the Hawking radiation. As density matrix of a system becomes more pure, its entropy becomes smaller (the entropy of a pure state is zero). 

In general, consider a composite system $AB = A\cup B$ divided into a bipartite division $(A,B)$ of region $A$ and region $B$.
When one member of an entangled pair is inside region $A$ and its partner is inside region $B$ there is a contribution to the entanglement entropy of the regions. In contrast, if both particles are either both inside region $A$, or  both inside region $B$,  then the pair contributes nothing to the entanglement entropy between the  two regions \cite{Cox_Forshaw:2022}. Put another way, if the composite state $AB$ of the system is given by a pure state $\rho_{AB}$, then entanglement entropy is given by the von Neumann entropy of either reduced density matrix 
$S(\rho_A) = S(\rho_B)$, and describes the bipartite entanglement \tit{between} region $A$ and region $B$.
Neither describes any possible entanglement \tit{within} (\tit{internal to}) the subsystem $A$, or within the subsystem $B$. As a trivial illustrative example, consider a  separable product of  two-mode squeezed vacuum states $\ket{\Psi}_{AB} = \ket{TMSV}_A\otimes\ket{TMSV}_B$ (two signal/idler pairs, each generated by SPDC \cite{Agarwal:2013,Gerry_Knight:2023}), 
one for system $A$ and one for system $B$, with composite density matrix $\rho_{AB} = \ket{\Psi}_{AB}\bra{\Psi}$. The reduced density matrix for subsystem $A$ is trivially given by $\rho_{A} = \Tr_B[\rho_{AB} ] = \ket{TMSV}_A\bra{TMSV}$, which is pure, so its von Neumann entanglement entropy is zero, even though the state itself  $\ket{TMSV}_A$ has a high degree of   bipartite entanglement between its \tit{internal} signal ($A_s$) and idler ($A_i$) modes (i.e. 
subsystem $A$ can be further decomposed internally as $A = A_s\cup A_i$, and similarly for subsystem $B$ ). The entanglement entropy $S(\rho_A) = S(\rho_B)$ only describes the entanglement between the regions (subsystems) $A$ and $B$ as a particular bipartite division of the composite system $AB$.

The net effect of the above concepts reproduces the Page curve of \Fig{fig:PageCurve}. For early times $t<t_{Page}$ (less than roughly half the Hawking radiation emitted), there is no island, and  
$S_R$ in \Eq{Cox:p249} is dominated by $S_{SC}$ which grows as Hawking's calculation indicated. 
At late times, $t>t_{Page}$, the island forms, and contributes with the QES 
very close to the horizon (see the  black dot at the right edge of the island in \Fig{fig:Cox:14.5:14.6}(left)), 
so that $Area(QES)/4\simeq Area(BH)/4 = \SBekHawk$.  
As the network of wormholes grows with the number of Hawking pairs generated, they reunite the Hawking particles internal to the horizon with their external partners outside the horizon, and act to purify the Hawking radiation, thus lowering to zero the contribution of $S_{SC}$ to $S_R$ in  \Eq{Cox:p249}. 
Thus, for $t>t_{Page}$, $S_R$ is dominated by the Area contribution in  \Eq{Cox:p249}, which itself is decreasing as the BH further evaporates, which forces $S_R$ to ``turn over." Again, this turnover point defines the Page time. (For further descriptive details see \cite{Almheiri_Hartman:2021,Cox_Forshaw:2022}, from which we have drawn, or the further technical details see  \cite{Penington:2020,Almheiri:2020}). 

At first glance, the use of the wormholes may seem like an ``accounting trick"  introduced to merely produce the Page curve \cite{Almheiri_Hartman:2021,Cox_Forshaw:2022}. But this is not the case. Rather it is seen now as a vindication of the RT formula, and the concept known as ``ER=EPR." Recall that the RT formula says the entanglement entropy between two regions $A$ and $B$ is associated with the area of the minimum area surface that divides the two regions. The ``ER=EPR" conjecture (Einstein-Rosen equals Einstein Podolsky Rosen) was put forth in 2013 by Maldacena and Susskind \cite{Maldacena_Susskind:2013} based on the idea that a spacelike curve connecting the Left and Right Rindler wedges of the maximally extended eternal Schwarzschild BH acts like a wormhole. This wormhole (Einstein-Rosen bridge) is only present if the composite bipartite quantum state of the  $L$ and $R$ Rindler wedges (i.e. the Minkowski vacuum) is the thermofield double state $\ket{TFD}$ (which in the case of harmonic oscillators is  the two-mode squeezed vacuum state $\ket{TMSV}$ of quantum optics). If there was no entanglement between the  
$L$ and $R$ Rindler wedges, there would be no wormhole. Said the other way, ``ER=EPR" conjecture claims entanglement ``creates" wormholes. With respect to the RT formula the question is where to draw the  minimal dividing surface. This is illustrated in \Fig{fig:Cox:14.5:14.6}(right). For $t<t_{Page}$ (upper right figure) when there are fewer interior Hawking particles and hence fewer wormholes connecting them to the outside, the RT surface is thought to cut through the wormhole (just as for the wormhole (ER-bridge) in the maximally extended eternal Schwarzschild Penrose diagram). However, for $t>t_{Page}$ (lower right figure), when more than half the Hawking radiation has been emitted, and there is a greater amount of entanglement between interior and exterior Hawking particles, there is a larger network of wormholes, and the RT surface becomes the QES at the edge of the island. This inside region of the BH is now illustrated
by the small black curve cutting the horizon in the lower right Penrose diagram in \Fig{fig:Cox:14.5:14.6}(right). It connects the QES at the end of the orange colored curve labeled B (island) behind the horizon, and  the orange colored curve marked A outside the horizon. The island is precisely that part of the interior that should be more correctly regarded as the outside of the BH.
The union of these two orange colored curves A and B is now considered as the outside of the BH, and contributes a net zero value to $S_{SC}$  contribution to the entanglement entropy  between the BH and the Hawking radiation R in \Eq{Cox:p249}, since the interior Hawking particles (now outside) act to purify the Hawking radiation. The entanglement slowly transfers from being between the BH and the radiation to being entirely within the Hawking radiation itself.

Far from being a purely descriptive account of BH evaporation, these concepts have also  been validated by computation means \cite{Almheiri_Hartman:2020, Penington:2022} through the use of Euclidean path integrals using the ``Replica Method." The Replica Method exploits the limit as $n\to 1$ of the formula for the Renyi entropy 
$S^{(n)}_{Renyi}(\rho) = (1-n)^{-1}\,\log\Tr[\rho^n]$ to reproduce the von Neumann entropy 
$S_{VNE}(\rho) = -\Tr[\rho\,\log \rho]$ (using L'Hopital's rule). One forms $n$ copies of $\rho$ on some spacelike slice of the spacetime and ``glues them together along the ``branch cut" bounded by the spacetime points $x$ and $y$ 
that defines the matrix elements
$\bra{\phi(x_f)} \rho \ket{\phi(x_i)}$ with respect to the initial and final state $\ket{\phi(x_i)}$ 
and $\ket{\phi(x_f)}$, respectively. This forms an $n-$fold Riemann sheet, with one copy of $\rho$ on each sheet.
An important aspect here is that one must also sum over all possible topologies, namely all possible ways of connecting the interiors of the BHs (one per sheet) behind the horizon. 
The  dominant contribution of the Euclidean path intregral  comes from its saddle points 
$e^{-(n-1)\,S(\rho_R)}$ for $n\simeq 1$, with  $S(\rho_R)$ given by \Eq{Cox:p249}, where $R$ is the region external to the BH horizon where the radiation escapes to.
The straightforward trivial stitching together of the $n$ copies of the replicated spacetime as $\rho^{\otimes n}$ (essentially $n$ copies of a single Riemann sheet) yields the Hawking contribution $S(\rho^{Hawking}_R)$. However, a non-trivial topology that cycles from the  $i$th sheet to the $(i+1)$st sheet as $i$ traverses from 1 to $n$ (as one circumnavigates the branch cut formed between the initial and final quantum states on each sheet) yields the replicated wormhole topology, where now island contribution to the extremum of $S(\rho_R)$ comes in to play. Minimizing over all contributions to the location of the island yields the final result
\be{Hartman:RWH:Notes:p20.4}
S(\rho_R) \simeq  \trm{min}\,
\left\{
\begin{tabular}{c}
$S(\rho_R)$, \\
$\quarter$\,Area($\pd$Island) + $S_{SC}$(Island $\cup$ R),
\end{tabular}
\right.
\ee
where $\pd$Island is the boundary of the Island region, interior to the BH horizon, and not necessarily connected to the external Hawking radiation region R by a spacelike slice through the spacetime (see the orange region $B$ in the lower right Penrose diagram in \Fig{fig:Cox:14.5:14.6}).
But this is precisely the statement of the Page curve, where the upper line of \Eq{Hartman:RWH:Notes:p20.4} is the Hawking result for for $t<t_{Page}$, the rising red-dashed curve in \Fig{fig:PageCurve}, and the lower line is falling black-dashed line for $t>t_{Page}$ where  
$\quarter\, Area(\pd \trm{Island})~\to~\quarter\,Area(BH)=\SBekHawk$.
(For details of the calculation see \cite{Almheiri_Hartman:2020,Penington:2022}; see also video Lecture 27 and associated class notes of Hartman's 2021 online course \tit{Black Holes and Quantum Information} \cite{Hartman_BHQI_course:2021}).

Interesting enough, while the Euclidean path integral techniques employed in the Replicated Wormhole method can directly calculate the entropy of the BH Hawking radiation, they do so without every knowing the state (density matrix) of the BH or the radiation R, or the composite pure state system (BH,R) explicitly.  It is an unproven, underlying assumption that $S(\rho_R)$ is related to some $\rho_R$ by the von Neumann entropy formula $S(\rho_R)~=~-\Tr[\rho_R \log \rho_R]$ (since the path integrals cannot directly compute the quantity $\log \rho_R$, hence the invocation of the Replica method). 

\subsection{The objectives of the explicit unitary phenomenological models of this present work with respect to the modern view of BH evaporation}\label{sec:ThisPaper}
In this section we present an overview of the motivations to the key features of the unitary phenomenological modes of this work that generalize the author's previous explorations in 
\cite{Alsing:2015,Alsing:2016}. Specific details will be presented in the following two sections.
\smallskip

The objectives of the unitary phenomenological models explored are to 
\begin{enumerate}[{(1)}]
\item \label{model:objective:1} Have the BH completely evaporate to zero mass, with the emitted Hawking radiation carrying away the total mass/energy of the BH,
\item \label{model:objective:2} Reproduce the key aspects of the Page curve (\Fig{fig:PageCurve}),
\item \label{model:objective:3} Account for the fate of the internal Hawking partner particles, especially by the end of the evaporation process, i.e. mimic the function of the replica wormholes without invoking wormholes.
\end{enumerate} 

These goals are achieved by using a fully quantized, trilinear Hamiltonian for SPDC (squeezed vacuum state) generation \cite{Agarwal:2013,Gerry_Knight:2023}, where the energy cost for signal/idler 
(exterior/interior Hawking particles) creation comes from the BH (pump). In this work, and addition new feature allows the generated idler/interior Hawking partner particles to further act as a SPDC pump sources for further signal/idler (Hawking pairs) generation. This process of  cascading sources of idler ``pumps" (waterfall) can then continue ad infinitum, and acts to transfer interior Hawking partner particles to the outside Hawking radiation, which acts to reduce the entanglement entropy.

The outline of the remainder of this paper is as follows.
 In  \Sec{sec:ModelReview}  we 
 review the unitary phenomenological models and results of \cite{Alsing:2015,Alsing:2016},  and the progress they obtained towards achieving the model objectives (\ref{model:objective:1}) and (\ref{model:objective:2}).
In \Sec{sec:BHWModel} we extend the previous two models to include the waterfall mechanism  to achieve model objective (\ref{model:objective:3}), and explore the Black Hole Waterfall (BHW) model for various ``depths" at which the interior idlers within the BH horizon can themselves act as a cascading set of additional pumps (SPDC sources). We present Page curves of the BH and Hawking radiation entropies as well as the Page Information and BH temperature.
In \Sec{sec:Summary} we summarize the results of the achievements and deficiencies of the previous two models \cite{Alsing:2015,Alsing:2016}, and discuss how the BH Waterfall model achieves all three of our stated model objectives.
In  \Sec{sec:Discussion} we discuss the qualitative similarity and difference between the BHW model and the modern view of BH evaporation discussed above. We also indicate future directions that can be explored with the BHW model.

\section{Review of the unitary phenomenological models and results of \cite{Alsing:2015,Alsing:2016}}\label{sec:ModelReview}
The models explored by the author in \cite{Alsing:2015,Alsing:2016}, motivated by related work by Adami and collaborators \cite{Adami_VerSteeg:2014,Bradler_Adami:2016},  are based on the following 
\tit{trilinear Hamiltonian} common in the quantum optics literature  for the generation of squeezed states by the process of spontaneous parametric down conversion (SPDC) \cite{Agarwal:2013,Gerry_Knight:2023}
 \be{H:eqn:BHW}
H_{p,i,s} = i\, r \left(
                         a_p a^\dagger_i a^\dagger_s 
                       - a^\dagger_p a_i a_s 
                  \right).
\ee
This model was also investigated by Nation and Blencowe \cite{Nation_Blencowe:2010} from a Heisenberg picture perspective, while \cite{Alsing:2015} explored this Hamiltonian from  a Schr\"{o}dinger picture.
Here the subscript $p$ denotes the ``pump," which here will be used to model the BH.
The subscripts $s$ and $i$ denote the generated correlated  ``signal" and ``idler" modes of the two-mode squeezed state created by the BH by the process of SPDC.
The signal and idler modes are taken to be created on opposite sides of the BH horizon 
at the expense of the energy of the BH.
We consider the signal mode $s$ as the particle that escapes to infinity as the Hawking radiation, and the idler mode $i$ as its entangled  partner particle behind the horizon.
SPDC  satisfy energy conservation $\om_p=\om_s+\om_i$, where  
$\om_p,\, \om_s,\, \om_i$ are the energy of the pump, signal and idler modes, respectively (using units, from now on, where $\hbar=c=1$). 

Squeezed state generation is natural to consider in the context of the Unruh effect \cite{Unruh:1976,Alsing_Milonni:2004} and Hawking radiation \cite{Hawking:1975} since it arises due to the exponential chirping of the  frequency observed by a uniformly accelerated stationary observer. This exponential chirp arises from the hyperbolic  nature of the Lorentz transformations required to transform a freely falling observer into the instantaneous rest frame of the uniformly accelerated observer \cite{Alsing_Milonni:2004}, giving rise to the celebrated Rindler transformation \cite{Rindler:1964,Rindler:1969}. Similarly, the exponential receding surface of an evaporating BH horizon can be modeled as a uniformly accelerated moving mirror, also giving rise to an exponential blue shift in the observed frequency \cite{Birrell_Davies:1982}. In fact, a simple exercise (see Problem 2.10, p47 of \cite{Agarwal:2013}, and associated references (26-30) therein) shows that if a harmonic oscillator  of frequency $\om_1$ initially in its vacuum ground state is abruptly changed to $\om_2$, then by the sudden approximation, the final state populations will be thermally distributed in a (single mode) squeezed vacuum state (see also \cite{Alsing_Dowling_Milburn:2005}).

\subsection{The model of \cite{Alsing:2015}}\label{subsec:review:Alsing:2015}
In the first model \cite{Alsing:2015}, the initial pure state of the composite system is given by 
\be{psi0:BHW}
\ket{\psi(0)} = \ket{\np0}_p\ket{0}_i \ket{n_{s0}}_s\defn  \ket{\np0}_p\ket{0}_{i,s},
\ee
and the general time dependent state satisfying the Schr\"{o}dinger equation 
$i\, \ket{\dot{\psi}(\tau)} = H_{p,i,s} \, \ket{\psi(\tau)}$
is given by
\be{psit:BHW}
\hspace{-0.1in}
\ket{\psi(\tau)} = \sum_{n=0}^{\np0} \, c_n(\tau)\, \ket{\np0-n}_p\ket{n}_i\ket{n_{s0}+n}_s
\defn\sum_{n=0}^{\np0} \, c_n(\tau)\, \ket{\np0-n}_p\ket{n}_{i,s}, 
\ee
leading to the coupled set of  ODE's for the (real) quantum amplitudes $c_n(t)$
\bea{cn:eqn:BHW}
\fl i \frac{d c_n (t)}{d t} &=& r \sqrt{n_{p0} - n} \, \sqrt{(n+1) (2\kappa + n)} \, c_{n+1}(t) \no
\fl                        &-& r \sqrt{(n_{p0} - n + 1)} \, \sqrt{n (2\kappa + n-1)} \, c_{n-1}(t),  \quad c_n(0) = \delta_{n,0}, \quad 2\kappa = n_{s0}+1.
\eea
Here, $n_{s0}$ (a ``seeded signal") accounts for any initial in-falling matter, but for simplicity, we will take $n_{s0}=0$ so that the initial state of the signal and idlers is their respective vacuum state 
$\ket{0}_i\ket{0}_s\equiv \ket{0}_{i,s}$. 
Note that while particle number is not conserved (since a signal/idler pair is created for every pump particle), total energy is conserved since from the occupation numbers in  \Eq{psit:BHW} we have
$(\np0-n) \om_p + n\om_s + n\om_i = \np0\om_p + n(-\om_p + \om_s + \om_i) = \np0\om_p$.
The model can easily be adapted for arbitrary initial states of the BH by pre-appending an additional summation over $\np0$, and modifying the quantum amplitudes, i.e.   
$\ket{\psi(\tau)}\to\ket{\Psi(\tau)}=\sum_{\np0=0}^{\infty}\sum_{n=0}^{\np0} \, c_{\np0,n}(\tau)\, \ket{\np0-n}_p\ket{n}_i\ket{n}_s$.

For physical BHs, a simple argument due to Zurek and Thorne \cite{Zurek_Thorne:1985} can be used to estimate $\np0$ for a BH of initial mass $M_0$. Let the BH be composed of $\np0$ particles of fixed mass $m_0$ of Compton wavelength $\frac{\hbar}{m_0 c}$ such that  $M_0 = \np0 m_0$. These collection of particles of mass $m_0$ cannot fit into a size smaller than the Schwarzschild radius $r_s = 2 G M/c^2$ of the BH without creating a BH themselves.
Thus,  setting $\frac{\hbar}{m_0 c}\sim \frac{2 G M_0}{c^2}$ yields $m_0 = \half M_{PL} (\frac{M_{PL}}{M_0})$. Setting $M_0 = \np0\,m_0$ yields $\np0 = \half (\frac{M_{PL}}{m_0})^2 \sim \half (\frac{10^{19} GeV}{m_0 c^2})^2$, or 
$(\frac{m_0}{M_{PL}})^2 = \frac{1}{2\,\np0}$. 
As the BH evaporates, we take its energy to be $E_{BH}=E_p(\tau) = \nbarp(\tau) m_0\,c^2 = M(\tau) c^2$, 
with $\nbarp(0)=\np0$.
The BH thermodynamic (coarse grained) entropy is then given by 
$\SBekHawk(\tau) = \quarter (4 \pi r^2_s) = 4\pi (\frac{M(\tau)}{M_{PL}})^2 = 4\pi \nbarp^2(\tau) (\frac{m_0}{M_{PL}})^2 = 2 \pi \frac{\nbarp^2(\tau)}{\np0}\sim E^2_{BH}$ 
%
\footnote{If one were to allow $m_0\to m(\tau)$ to vary with the changing mass of the BH $M(\tau) = \nbarp(\tau)\,m(\tau)$, an equivalent calculation yields $(\frac{m(\tau)}{M_{PL}})^2 = \frac{1}{2\,\nbarp(\tau)}$, so that
$E_{BH}(\tau)=\nbarp(\tau)\,m(\tau) = \frac{1}{\sqrt{2}}\,\sqrt{\nbarp(\tau)}\,M_{PL}$, which yields 
$\SBekHawk(\tau) = 2\,\pi\,\nbarp(\tau)$, so that once again, $\SBekHawk(\tau)\sim E^2_p(\tau)$. 
For simplicity, in this work we will follow the scaling in the main text with fixed mass $m_0$, and measure the BH energy as $E_{BH}(\tau)/m_0 = \nbarp(\tau)$.},
while the fine-grained von Neumann entropy is given by $S(\rho_{BH})~=~S(\rho_p)~\equiv~S(\rho_{i,s})$ 
(with the last equality following since the composite state is pure).

Crucial to this model is that we take  $\np0$ as a large, but \tit{finite} number, and consequently keep track of the upper limit on the sum over $n$ to be $\np0$. The rationale is that this then models the evaporation of the  pump/BH, since creation of signal/idler Hawking pairs depletes the energy of the BH, via the term 
$\ket{\np0-n}_p$ in $\ket{\psi(\tau)}$. Importantly, for early times in the evolution, when the average number of particles in the pump is much greater than number of emitted particles $\nbarp \gg \nbars, \nbari$, 
(where $\nbarp = \bra{\psi(\tau)} a^\dag_p a_p\ket{\psi(\tau)}$, etc\ldots)
one can approximate $\ket{\np0-n}_p\simeq \ket{\np0}_p$ and factor it out of the sum over $n$ to obtain the approximate state 
\bea{psi:shorttime:BHW}
\fl \ket{\psi_<(z,\tau)} &\approx& \ket{\np0}_p\,\ket{\phi}_{i,s}, \quad z= \tanh^2(\tau) < z^* \approx 0.506407 \hspace{0.5em} \textrm{for} \hspace{0.5em} \np0 \gg 1,\; \tau = \sqrt{\np0}\,r\,t, \label{psi:shorttime:BHW:1}\\
%
\fl \ket{\phi}_{i,s}^{(sqzd)} &=& \sqrt{\frac{1-z}{1-z^{\np0+1}}}\,
                         \sum_{n=0}^{\np0} z^{n/2} \, \ket{n}_{i,s} \approx
 \sqrt{1-z}\,\sum_{n=0}^{\infty}\, z^{n/2} \ket{n}_{i,s}, \; (\np0\rightarrow\infty),  \label{psi:shorttime:BHW:2}
\eea
where $\ket{n}_{i,s} = (a^\dag_i\,\asdag)^n/n!\,\ket{0}_{i,s}$.
This is what we expect for early times in the BH evaporation evolution. In the limit $\np0\to\infty$ the state  
$\ket{\phi}_{i,s}^{(sqzd)}$ in \Eq{psi:shorttime:BHW:1} is the well known entangled two-mode squeezed vacuum state $\ket{TMSV}$ \cite{Agarwal:2013,Gerry_Knight:2023} with probabilities $p_n = (1-z)\,z^n$. 
In this case the Hamiltonian $H_{p,i,s}$ in \Eq{H:eqn:BHW} is being approximated by a constant amplitude 
$\xi$, classical, non-depleting pump $H_{p,i,s}\to H_{\xi,i,s} =  i\,\xi (a^\dagger_i a^\dagger_s-a_i a_s )$ 
known as the two-mode squeezing Hamiltonian in quantum optics.

For a harmonic oscillator with energy $E_n = n \hbar\om$ (dropping the zero point energy $\thalf\hbar\om)$, this is the same as the thermofield double state $\ket{TFD}$ discussed previously. In the case of the maximally extended Schwarzschild spacetime (or in flat Minkowski spacetime for the Unruh effect \cite{Unruh:1976}) we can consider $\ket{\phi}_{i,s}^{(sqzd)}\to\ket{TFD}$ as a representation of the Minkowski vacuum $\ket{0}_M$ in terms of Rindler modes, with the signal mode $\ket{n}_s\to\ket{n}_R$ belonging to the Right Rinder Wedge, and the idler mode 
$\ket{n}_i\to\ket{n}_L$ belonging to the Left Rinder Wedge \cite{Unruh:1976,Carroll:2004}.
For finite $\np0$ we define the normalized probability $p_n(z;\np0)$ to be
\be{pn:shorttime:BHW}
p_n^<(z) = \frac{(1-z)}{(1-z^{\np0+1})} \, z^n, \qquad  \sum_{n=0}^{\np0} p_n^<(z) =1.
\ee
The two-mode squeezed state has the well know property that if we trace over the idler modes, the reduced density matrix for the signal is thermal 
$\rho^{(sqzd)}_s~=~\Tr_i[\ket{\phi}_{i,s}^{(sqzd)}\bra{\phi}_{i,s}^{(sqzd)}] = (1-z)\sum_{n=0}^\infty z^n\ket{n}_n\bra{n}$ with $z\leftrightarrow e^{-\beta\,\omega}$, $\beta = 1/T$.

The salient point of this model is that at later times when $\nbarp \approx \nbars, \nbari$, one can no longer factor out the state $\ket{\np0}_p$ from the sum, and entanglement now builds up between the pump and the signal/idler modes, (BH and exterior/interior Hawking particles). This was explored in \cite{Alsing:2015} for both early and late time evolutions, and for entanglement properties. For this model it is easy to see that the all the reduced density matrices are diagonal and their von Neumann entropies are identical, e.g.
$\rho_p = \Tr_{i,s}[\ket{\psi(\tau)}\bra{\psi(\tau)}]= \sum_{n=0}^{\np0} |c_n(\tau)|^2 \ket{\np0-n}_p\bra{\np0-n}$,
$\rho_{s} = \Tr_{p,i}[\ket{\psi(\tau)}\bra{\psi(\tau)}]= \sum_{n=0}^{\np0} |c_n(\tau)|^2 \ket{n}_s\bra{n}, \ldots$
with $S(\rho_p)\equiv S(\rho_{i,s}) = S(\rho_{s})=S(\rho_{i})$, since they all contain the same probabilities 
$p_n(\tau) = |c_n(\tau)|^2$ (the first equality automatically follows since the composite state is pure).
\begin{figure}[h]
\begin{center}
\begin{tabular}{ccc}
\includegraphics[width=2.5in,height=1.75in]{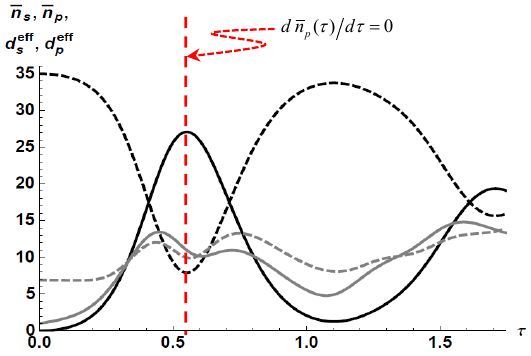} &
{} &
\includegraphics[width=2.75in,height=1.75in]{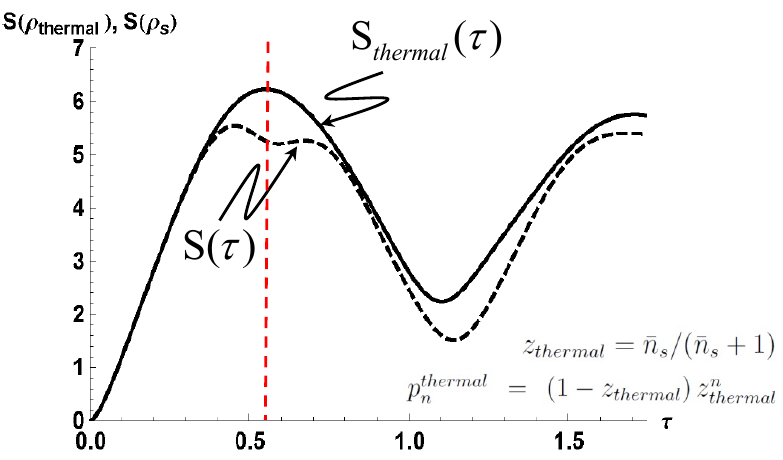}
\end{tabular}
\end{center}
\caption{(left) Evolution of the average number or particles in the pump/BH  $\nbarp(\tau)$ (black dashed line), and the signal/Hawking radiation $\nbars(\tau)$ (black solid) for the BH in an initial coherent state with $\nbarp(0)$ = 35, and the signal/idlers in the vacuum state.
$d^{eff}_p \defn 1 + \Delta\nbarp$ (gray dashed) and 
$d^{eff}_s \defn 1 + \Delta\nbars$ (gray solid) are the effective dimensions of the pump/BH and signal/Hawking radiation in terms of the variance in their particle number \cite{Alsing:2015,Nation_Blencowe:2010}.
Vertical red-dashed line indicates the time at which 
$d\nbarp(\tau)/d\tau =0$ again for $\tau>0$, the limit of the validity of the model.
(right) von Neumann entropies  for: 
$S(\rho_s)$ signal/Hawking radiation (dashed-black curve),  
$S(\rho_{thermal})$ effective thermal state  (solid black curve) formed from $\rho_s$.
The difference between the solid-black and dashed-black curve is the Page Infomation
$I(\tau) \defn S(\rho_{thermal})- S(\rho_s)$.
}\label{fig:from:Alsing2015}
\end{figure}

In \Fig{fig:from:Alsing2015}(left) we plot the evolution of the average number or particles $\nbarp(\tau)$  in the pump/BH  (black dashed curve), and in the signal/Hawking radiation $\nbars(\tau)$ (black solid), for the BH in an initial coherent state  with $\nbarp(0) = 35 = \alpha^2$ 
\footnote{The plots of $\nbarp(\tau)$ and $\nbars(\tau)$ have a similar form for any value of $\np0$ as long as it significantly different than the initial number of particles in the signal/idler modes, which we have taken to be zero. Plots using large initial values of $\np0$ simply change the scale on the $\tau$ axis.},
and the signal/idlers in the vacuum state, 
$\ket{\psi(0)} = \ket{\alpha}_p\ket{0}_i\ket{0}_s$, with  
$\ket{\alpha}_p = e^{-\alpha^2/2}\sum_{m=0}^\infty \frac{\alpha^m} {\sqrt{m!}}\ket{m}_p$ \cite{Agarwal:2013, Gerry_Knight:2023}. For early times both $\nbarp(\tau)$ and $\nbars(\tau)$ have nearly zero slope, and for these time with $\nbarp(\tau)\gg \nbars(\tau)$ the Hawking radiation is essentially composed of squeezed signal/idler pairs, and hence is thermal. We define an effective thermal state from $\rho_s$ by defining 
$\rho_{thermal}= \sum_{m=0}^\infty \frac{\nbars^m(\tau)}{(\nbars(\tau)+1)^{m+1}}\ket{m}_s\bra{m}$. 
The Page Information \cite{Nation_Blencowe:2010,Alsing:2015} $I(\tau) \defn S(\rho_{thermal})- S(\rho_s)$ represents the information that leaks out of the BH due to the correlations (entanglement) built up between the pump/BH and signal/Hawking radiation as the system evolves. This is shown in  \Fig{fig:from:Alsing2015}(right).

Note that the validity of this model should really only be considered up to the time when $d\nbarp(\tau)/d\tau=0$ again for $\tau>0$, depicted as the vertical red-dashed line in \Fig{fig:from:Alsing2015}. After this time the signal/idler start ``acting" as the effective ``pump," and the population in the pump/BH 
increases again.\footnote{In fact for very long times the populations continue to oscillate and eventually  equilibrate, as explored in \cite{Birrittella:2020}, with the pump rapidly depleting by roughly $80\%$ of its initial value, a phenomena known in the quantum optics community, see  \cite{Bandilla:2000}}. 
When $\nbarp \gg \nbars,\nbari$, the term $a_p a^\dag_i a^\dag_s$ dominates the evolution. However, since the Hamiltonian is Hermitian, the conjugate term $a^\dag_p a_i a_s$ must be present, and is still acting on the system. It is only when $\nbarp \approx \nbars,\nbari$ that this conjugate term truly ``kicks in" and the role of ``pump" (the dominant ``driver" of SPDC) switches from the BH to the Hawking pairs. Physically, this could only happen if the whole system of the BH/radiation was in some sort of cavity. 
It is interesting that the point at which $d\nbarp(\tau)/d\tau=0$ again is later than the time when 
$\nbarp =\nbars$. There is a time delay after the populations swap for the signal/idlers (Hawking radiation) to now acts the effective pump, driving the BH.

The One Shot model of \cite{Alsing:2016}, discussed in the next section, was constructed to rectify this latter point above, by allowing the BH to act for a short time creating a signal/idler pair, with the signal escaping to infinity as Hawking radiation. Afterwards, the BH acts on a different vacuum mode, creating a new (different) signal/idler pair, with the signal again, escaping to infinity, and the process is repeated ad infinitum.

There still remains a final energy issue with this first model. Even if  we were able to have the pump/BH complete evaporate so that the final state was given by $\ket{0}_p\ket{\np0}_i\ket{\np0}_s$ with $n\to\np0$ in \Eq{psit:BHW}, only half the energy $\np0 \om_p/2$ of the BH has been carried away 
by the exterior signal/Hawking radiation 
$\ket{\np0}_s$ (assuming for simplicity, and without loss of generality, degenerate SPDC with 
$\om_s=\om_i=\om_p/2$), and the other half remains behind the horizon as the interior Hawking particles. So this model \cite{Alsing:2015} appears to leave an undesired BH remnant at the end of the evaporation process. Something akin to the "ER=EPR" wormhole mechanism appears to be needed to avoid this situation. 

We will see in the third model considered in  \Sec{sec:BHWModel}, constructed in order to address model objective (3), this issue can also be rectified by assuming that the interior Hawking particles can also create their own interior/exterior idler/signal Hawking pairs, now with half of each interior particle's energy, i.e. $\om_p/4$. With each newly created idler acting as its own ``pump", this cascading (waterfall) process can be carried out ad infinitum so that the total energy carried away by the collection of all the  signals/Hawking radiation is
$\np0\, \om_p\, \sum_{d=1}^{D} (\half)^d  = (1 - (\half)^D)\,\np0\,\om_p 
\mylimit{{D\to\infty}}{=}
\np0\,\om_p
$, 
the initial mass of the BH 
(this  ``remnant mitigation" proposal was first suggested by the author in \cite{Alsing:Taiwan:Talk:RQI:2019}). 
What remains inside the horizon are $\np0$ idler particles of arbitrarily small vanishing energy 
$(\half)^D\,\np0\,\om_p 
\mylimit{{D\to\infty}}{=}
0$.
This phenomena of zero energy remaining in the internal (``soft-hair") horizon modes of the BH was explored in a 2018 paper by Hotta, Nambu and Yamaguchi entitled \tit{Soft-hair-enhanced entanglement beyond the Page Curves in Black Hole evaporation Qubit model} \cite{Hotta_Nambu_Yamaguchi:2018}. In this work, an infinite number of super-translation charges appear to be able to store the whole quantum information of the absorbed matter forming the BH, arising from a near-horizon symmetry providing a massive degeneracy of the Hamiltonian (the approximately zero energy ``soft hair" of the BH horizon).
The impetuous for the BH Waterfall Model presented here in this work was inspired by this work, with a desire to develop a qualitatively similar model, but from a squeezing Hamiltonian perspective. 
Before we explore this model in \Sec{sec:BHWModel}, which we descriptively denote as the 
\tit{Black Hole Waterfall}, we first review  the second One Shot model \cite{Alsing:2016} in the next section, on which the former is built. 

\subsection{The One Shot model of \cite{Alsing:2016}}\label{subsec:OneShot:2016}
The second model \cite{Alsing:2016} applied the One Shot prescription of Br\'{a}dler and Adami \cite{Bradler_Adami:2016} to the first model \cite{Alsing:2015}. 
The second model begins with the initial state
\be{Psi:BA:BHW}
\ket{\Psi(0)} = \ket{\np0}_p\prod_{k'=1}^N \ket{0}_{i_{k'},s_{k'}}=
\ket{\np0}_p\otimes
\ket{0}_{i_1,s_1}\otimes
\ket{0}_{i_2, s_2}\otimes\ldots\otimes
\ket{0}_{i_N, s_N},\quad
\ee 
where $\tau = N \Delta\tau$, and $N$ is the number of time slices.
The evolution of the state $\ket{\psi(0)}$ in \Eq{Psi:BA:BHW} is given by \cite{Bradler_Adami:2016}
\be{U:T:BHW}
\hspace{-0.1in}
\fl \ket{\psi(\tau)} = U(\tau,0) \ket{\Psi(0)}
= \mathcal{T} e^{-i\int d\tau' \, H_{p,s,i}(\tau')}\,\ket{\Psi(0)}
\approx \prod_{k=1}^N \,
e^{-i H_{p,i_k,s_k}\Delta\tau}\,\ket{\np0}_p\,
\prod_{k'=1}^N\,\ket{0}_{i_{k'},s_{k'}},\;
\ee
where ${\mathcal{T}}$ is the time-ordered product and in the second equality we have used a simplified version of the Trotter expansion valid for $N$ small time
slices of size $\Delta\tau$, with $U_{p,k} = e^{-i H_{p,i_k,s_k}\Delta\tau}$ acting on modes $p$ and $(i_k,s_k)$.

After the first time slice, the wave function is
\bea{Psi:1:BHW}
\fl \ket{\Psi(1)} &=&U_{p,1}\,\ket{\Psi(0)} =  \sum_{n_1=0}^\np0 \sqrt{p_{n_1}^{(n)}(z)}\,\,
     \ket{\np0-n_1}_p\ket{n_1}_{i_1,s_1}\otimes
     \prod_{k'=2}^N\,\ket{0}_{i_{k'},s_{k'}}, \qquad z \ll z^*, \\ \label{Psi:1:exact}
\fl  &\equiv& \sum_{n_1=0}^\np0 \sqrt{p_{n_1}^{(\np0)}(z)}\,\,
     \ket{\np0-n_1}_p\ket{n_1}_{1},
     \; p_{n_1}^{(\np0)}(z) = \frac{(1-z)}{(1-z^{\np0+1})}\,z^{n_1},
     \; \sum_{n_1=0}^\np0\,p_{n_1}^{(\np0)}=1,\qquad \label{Psi:1:exact:probs}
     \\
\fl &\approx&
\ket{\np0}_p\otimes\sum_{n_1=0}^\np0 \sqrt{p_{n_1}^{(\np0)}(z)} \,\, \ket{n_1}_{1},     \qquad \np0\gg n_1, \\
\fl &\equiv&
\ket{\np0}_p\otimes \ket{\phi^{(sqzd)}}_{1},\label{Psi:1:approx}
\eea
where $\ket{\phi^{(sqzd)}}_{1}=(1-z)\sum_{n_1=0}^{\np0\rightarrow\infty} z^{n_1} \ket{n_1}_1$ 
is two-mode
signal/idler emittted Hawking radiation state.
The emitted Hawking signal/idler pairs are approximately squeezed for early time $z<z^*$, due the approximate factorization of the pump/BH form the generated signal/idler states.
However, for long time evolution the exact state in \Eq{Psi:1:exact} does not factorize as in case of the
short time state \Eq{Psi:1:approx}, and correlations and entanglement build up between the pump/BH and the interior/exterior Hawking pairs.
Note that the notation $p_{n_1}^{(\np0)}$ indicates the probability that $n_1$ particles are emitted into the Hawking radiation signal/idler mode when there were initially $\np0$ particles in the BH `pump' mode.
Henceforth, we shall
denote  $\ket{n_k}_{k}\equiv \ket{n_k}_{i_k, s_k}$, drop the argument $z$ on the probabilities,
and leave implied the unoccupied vacuum signal/idler states $\ket{0}_{i_{k'},s_{k'}}$ for $k'$ greater than the current timeslice considered. 
From \Eq{Psi:1:BHW} and \Eq{pn:shorttime:BHW} 
the state $\ket{\Psi(1)}$ is clearly normalized to unity.

To illustrate the notation employed, it is instructive to write down the
wavefunction at after the second emission event
\bea{Psi:2:BHW}
\fl \ket{\Psi(2)} &=& U_{p,2}\,\ket{\Psi(1)}
              = \sum_{n_1=0}^\np0 \, \sum_{n_2=0}^{\np0-n_1}\,
              \sqrt{p_{n_1}^{(\np0)}\,p_{n_2}^{(\np0-n_1)}}\,
                 \ket{(\np0-n_1)-n_2}_p\,\ket{n_1}_{1}
                                         \,\ket{n_2}_{2}, \\ \label{Psi:2:exact}
\fl &\approx&
 \ket{\np0}_p\otimes
\sum_{n_1=0}^\np0 \,  \sqrt{p_{n_1}^{(\np0)}}\,\ket{n_1}_{1}\otimes
\sum_{n_2=0}^{\np0-n_1}\,\sqrt{p_{n_2}^{(\np0-n_1)}}\,\ket{n_2}_{2}, \quad \np0 \gg n_1, n_2,\no
\fl &\approx&\ket{\np0}_p\otimes\ket{\phi^{(sqzd)}}_1\otimes\ket{\phi^{(sqzd)}}_2, \label{Psi:2:BHW:2}
\eea
where
\be{Psi:2:approx:BHW}
p_{n_2}^{(\np0-n_1)}(z) = \frac{(1-z)}{(1-z^{(\np0-n_1)+1})}\,z^{n_2},
     \qquad \sum_{n_2=0}^{\np0-n_1}\,p_{n_2}^{(\np0-n_1)}=1. 
\ee
The new feature of \Eq{Psi:2:BHW} is that the second particle has been emitted into the second signal/idler mode with the only dependence upon mode $1$ being that the initial number of particles in the BH `pump' source is now $\np0-n_1$, where $n_1$ is the number of particles that were emitted into mode $1$ during the first emission event
(note: $n_1\in[1,\np0]$).
Again, in the short time limit \Eq{Psi:2:BHW:2} indicates that the emitted Hawking radiation is approximately a succession of independent two-mode squeezed states in modes $1$ and $2$ respectively.

Note that by utilizing a wavefunction $\ket{\Psi(2)}$ we are implicitly assuming a degree of coherency between the pump and the emitted Hawking radiation signal/idler modes, as exhibited in the exact states for $\ket{\Psi(1)}$ and $\ket{\Psi(2)}$ in \Eq{Psi:1:BHW} and \Eq{Psi:2:BHW} respectively.
One regime of the One Shot decoupling procedure  is to approximately decouple the emitted Hawking radiation modes from the pump at each emission event, while also keeping track of the finite and decreasing nature of the BH quantized degree of freedom $\nbarp(\tau)$ that arises from the finite, though large, initial occupation number 
$\np0\gg 1$. In the language of laboratory SPDC, on is making the implicit assumption that the coherency of the BH `pump' source is shorter than the average time between emission events. 
However, one can also consider another regime which assumes a longer coherence time between the pump/BH emission events, allowing for stronger correlations and entanglement to develop. Depending on the assumed coherence time of the pump/BH (``laser") relative to the average time between signal/idler emission events, correlations between signal/idler pairs emitted by the BH at different times can occur, since such pairs are indirectly coupled to each other through their separate interactions with the pump (see the Discussion section of \cite{Alsing:2016}).

Since each unitary emission $\{U_{p,i}\}_{i=1:N}$ acts for a short time $\Delta\tau$,
we are continually in the short time regime
$z<z^*$ and each emitted signal/idler Hawking radiation pair is nearly, but not exactly, a two-mode squeezed state. However, the occupation number of the BH `pump' mode is continually decreasing, and it is the effect of this finite nature of the `pump' source on the total state that we wish to examine for long times (large $N$) as the BH evaporates.
Consider the wavefunction $\ket{\Psi(N)}$ after $N$ emitted events given by the generalization of \Eq{Psi:2:BHW}
\bea{Psi:N:BHW}
\fl \ket{\Psi(N)}
&=&
       \sum_{n_1=0}^\np0 \,
       \sum_{n_2=0}^{\np0-n_1}\,
       \sum_{n_3=0}^{\np0-(n_1+n_2)}\ldots\sum_{n_N=0}^{\np0-(n_1+\ldots+n_{N-1})}\,
        \sqrt{p^{(n)}_{n_1}\,p_{n_2}^{(\np0-n_1)}\,p_{n_3}^{(\np0-n_1-n_2)}\ldots
        p_{n_N}^{(\np0-n_1-\ldots-n_{N-1})}}\,\no
\fl &\times&
 \ket{\np0-(n_1+\ldots+n_N)}_p
 \otimes\prod_{i=1}^N \,\ket{n_i}_{i}, \\
\fl &\approx&
  \ket{\np0}_p\otimes
  \prod_{i=1}^N \,\ket{\phi^{(sqzd)}}_{i}, \qquad \np0 \gg \left.\{n_i\}\right|_{i=1:N}\label{Psi:N:approx}, \\
\fl &\equiv&
  \ket{\np0}_p\otimes\ket{\Phi^{(sqzd)}(N)}.
\eea
By construction we have $\bra{\Psi(N)}\Psi(N)\rangle=1$.

Let us rewrite $\ket{\Psi(N)}$ as follows. We define $j_i = \sum_{m=0}^{i} n_m$ with $j_0\equiv0$. Keeping track of the upper and lower limits on each summation, we obtain the representation
\bea{Psi:j:jN:last:BHW}
\fl \ket{\Psi(N)}
&=&
       (1-z)^{N/2}\,\sum_{j_1=0}^\np0 \,
       \sum_{j_2=j_1}^{\np0}\,
       \sum_{j_3=j_2}^{\np0}
       \ldots
       \sum_{j_N=0}^{\np0}\,
       \sqrt{z^{j_N}} \ket{\np0-j_N}_N \otimes
       \prod_{i=1}^N \frac{1}{\sqrt{(1-z^{\np0-j_i}+1)}} \, \ket{j_i-j_{i-1}}_i, \no
\fl &=&
       (1-z)^{N/2}\
       \sum_{j_N=0}^{\np0}  \,\sqrt{z^{j_N}} \, \ket{\np0-j_N}_N\,\otimes \no
\fl & &
 \hspace{0.75in} \left[
       \sum_{j_1=0}^{j_N} \,
       \sum_{j_2=j_1}^{j_N}\,
       \sum_{j_3=j_2}^{j_N}
       \ldots
       \sum_{j_{N-1}=j_{N-2}}^{j_N}\,
       \prod_{i=1}^N \frac{1}{\sqrt{(1-z^{\np0-j_{i-1}+1})}} \, \ket{j_i-j_{i-1}}_i,
 \right] \qquad \label{Psi:j:jN:first:BHW} \\
\fl &\equiv&
(1-z)^{N/2}\,
\sum_{j_{N}=0}^{\np0}\,  \,\sqrt{z^{j_N}} \, \ket{\np0-j_N}_N\otimes\ket{\tilde{\Phi}^{'(N)}_{j_N}},   \label{Psi:N:BHW}
\eea
where we have defined the unnormalized state $\ket{\tilde{\Phi}^{'(N)}_{j_N}}$ by the expression in the large square brackets in
\Eq{Psi:j:jN:first:BHW}, and we have pulled the sum over the collective pump/BH emission index $j_N$ to the far left, which alters the limits of the remaining inner nested sums.
$\ket{\tilde{\Phi}^{'(N)}_{j_N}}$ describes the emitted Hawking radiation state with exactly $j_N$ particles
(at the $N$th time slice) emitted into $N$ possible distinct signal/idler modes,
which is, in general, a superposition state over all Fock states 
whose occupation numbers sum to exactly $j_N$.

In \cite{Alsing:2016}  the term $(1-z^{\np0-j_{i-1}+1})^{-1/2}$ inside $\ket{\tilde{\Phi}^{'(N)}_{j_N}}$ was approximated in two ways. First, since we take $z<1$ and $\np0\gg 1$, a zeroth order approximation is to simply set this factor to unity for all $j_{i-1}$.
This defines a normalized state 
\bea{Phi:N:approx:BHW}
\ket{\Phi^{'(N)}_{j_N}} &=&
\frac{1}{\tiny \sqrt{\left(
           \begin{array}{c}
             j_N+N-1 \\
             j_N
           \end{array}
\right)}}\,
       \sum_{j_1=0}^{j_N} \,
       \sum_{j_2=j_1}^{j_N}\,
       \sum_{j_3=j_2}^{j_N}
       \ldots
       \sum_{j_{N-1}=j_{N-2}}^{j_N}\,
       \prod_{i=1}^N 
       \, \ket{j_i-j_{i-1}}_i,\no
&=&
\frac{1}{\tiny \sqrt{\left(
           \begin{array}{c}
             j_N+N-1 \\
             j_N
           \end{array}
\right)}}\,
\sum_{j_1\le j_2\ldots \le j_{N-2}\le j_{N-1}}^{j_N}\,\,
\prod_{i=1}^N \, \ket{j_i-j_{i-1}}_i,
\eea
where the binomial factor 
$
{\tiny \left(
           \begin{array}{c}
             j_N+N-1 \\
             j_N
           \end{array}
\right)}
$
in \Eq{Phi:N:approx:BHW} counts the number states containing exactly $j_N$ Hawking radiation particles into $N$ signal/idler modes,
i.e. the selection of $j_N+N-1$ objects taken $j_N$ at a time \textit{with} repetitions.
We can also intuitively understand the nested sum in \Eq{Phi:N:approx:BHW}
over the dummy indices $j_1\le j_2\ldots \le j_{N-2}\le j_{N-1}$  as the number of lattice points in the `upper diagonal' quadrant (including the diagonal)
of a $N-1$ dimension hypercube with $j_N+1$ lattices points ($0,1,\ldots,j_N$) per dimension.

As an example, the normalized signals/idlers state $\ket{\Phi^{'(N=4)}_{j_N=2}}$ 
of exactly two particles emitted by the BH ($j_N=2$) after $N=4$ emission events  is given by
\bea{Phi:N4:jN2:BHW} 
\fl \ket{\Phi^{'(N=4)}_{j_N=2}}
 &=&
\left(\ket{2,0,0,0} + \ket{0,2,0,0} +\ket{0,0,2,0} +\ket{0,0,0,2} +\ket{1,1,0,0}
\right. \no
\fl &+&
\left.
 \ket{1,0,1,0} +\ket{1,0,0,1} +\ket{0,1,1,0} +\ket{0,1,0,1} + \ket{0,0,1,1}
\right)_{1,2,3,4}/\sqrt{10}, \\
\fl & & \textrm{with the number of component states:} \quad  
\left.\left(
           \begin{array}{c}
             j_N+N-1 \\
             j_N
           \end{array}
\right)\right|_{N=4,j_N=2} \hspace{-.5in}= 10. \nonumber
\eea
Note that the state $\ket{\Phi^{'(N=4)}_{j_N=2}}$ is a highly entangled state  between its component signal and idler states, exhibiting entanglement between different emission times. For example, tracing out over the signal/idler pairs $2$ and $3$ 
yields a reduced density matrix  
$\sigma_{1,4} ~=~\Tr_{2,3}[\ket{\Phi^{'(N=4)}_{j_N=2}} \bra{\Phi^{'(N=4)}_{j_N=2}}]$
for signal/idler pairs $1$ and $4$ emitted at $N=1$ and $N=4$, containing up to two excitations
\footnote{Recall $\ket{nm}_{14} = \ket{n}_1\ket{m}_4 \equiv \ket{n,n}_{i_1,s_1}\ket{m,m}_{i_4,s_4}$.}
\bea{sigma14:BHW}
\hspace{-0.5in}
\sigma_{1,4} &\equiv&  
q_0 \ket{\varphi_0}_{1,4}\bra{\varphi_0} +
q_1 \ket{\varphi_1}_{1,4}\bra{\varphi_1}  + 
q_2 \ket{\varphi_2}_{1,4}\bra{\varphi_2}, \no
\hspace{-0.5in}
 \ket{\varphi_0}_{1,4}   &=& \ket{00}_{1,4}, \qquad
\ket{\varphi_1}_{1,4}     =  \tfrac{1}{\sqrt{2}}\,(\ket{01}_{1,4} + \ket{10}_{1,4}), \no
\hspace{-0.5in}
\ket{\varphi_2}_{1,4}   &=& \tfrac{1}{\sqrt{2}}\,(\ket{02}_{1,4} + \ket{11}_{1,4} +\ket{20}_{1,4} ),
\quad (q_0,q_1,q_2) = (3,4,3)/10, \no
\hspace{-0.5in}
S(\sigma_{1,4}) &=& -\sum_{i=0}^{2} q_i\,\log q_i.
\eea
A further trace over $i_1,i_4$ yields the $s_1,s_4$ density matrix
\bea{sigma:s1s4:BHW}
\hspace{-1.5in}
\sigma_{s_1,s_4} &\equiv&  
\frac{1}{10}
\Big[
3 \ket{00}_{s_1,s_4}\bra{00} + 
2 \left(
\ket{01}_{s_1,s_4}\bra{01} + \ket{10}_{s_1,s_4}\bra{10}
\right)
+ \left(\ket{02}_{s_1,s_4}\bra{02} + \ket{11}_{s_1,s_4}\bra{11} + \ket{20}_{s_1,s_4}\bra{20}\right)
\Big]. \qquad
\eea

Continuing, upon replacing $j_N\rightarrow k$ (to simplify notation) as the total number of Hawking radiation particles emitted into $N$ signal/idler modes, we obtain the first approximation
\bea{Psi:Normalized:BHW}
\ket{\Psi(N)}
&\approx&
\sum_{k=0}^{\np0} \sqrt{P_{k}^{'(N)}}\, \,\ket{\np0-k}_p \,\ket{\Phi'^{(N)}_{k}},  \label{Psi:Normalized:BHW:1} \\
\;\;
P_{k}^{'(N)} &=& \frac{\tilde{P}_{k}^{'(N)}}{\sum_{k'=0}^{\np0}\,\tilde{P}_{k'}^{'(N)}},  \quad
\tilde{P}_{k}^{'(N)} = (1-z)^N\, z^{k}\,
\left(
       \begin{array}{c}
          k + N-1 \\
          k
        \end{array}
      \right).  \label{Psi:Normalized:BHW:2}
\eea
%
Note that in the limit $\np0\rightarrow\infty$ we have $\sum_{k=0}^{\infty}\,\tilde{P}_{k}^{(N)}=1$ using the identity
$
\sum_{k=0}^{\infty}\,z^k
\tiny{
\left(
       \begin{array}{c}
          k + N-1 \\
          k
        \end{array}
      \right)
}
= (1-z)^{-N}.
$
For finite $\np0$ we normalize the probabilities, as in \Eq{Psi:Normalized:BHW:2}.

It turned out that this first approximation turns over the entropy curve of the BH, 
brings it to zero asymptotically only for for very long times.
A second, finer approximation was made as follows.
Recalling that $j_0=0$, we can factor out from all the nested summations an overall constant term $(1-z^{\np0+1})^{-1/2}\rightarrow 1$ for $z\ll 1$ and any reasonable sized value of $\np0$. The remaining factors have to be summed from $j_i\in[j_{i-1},j_N\equiv k]$, in succession from the inner summations, outwards. These complicated nested sums are what led to the numerical lattice-path approach of Br\'adler and Adami \cite{Bradler_Adami:2016}. Here, we make the simplified, but reasonable approximation that $(1-z^{\np0-j_{i-1}+1})^{-1/2}$, is dominated by its largest contribution $j_i=k$ from the upper limit of the summation, yielding $(1-z^{\np0-k+1})^{-1/2}$ which can then be factored out of all the nested summations, except the outermost one over $k$ itstelf. Since there are $N-1$ inner summations at time $N$ we obtain $\ket{\Psi(N)}$  with the slightly refined second approximation to the probabilities
\bea{Pksi:corrected:BHW}
\hspace{-0.75in}
 \ket{\Psi(N)}
&\approx&
\sum_{k=0}^{\np0} \sqrt{P_{k}^{(N)}}\, \,\ket{\np0-k}_p \,\ket{\Phi^{(N)}_{k}}, \label{Pksi:corrected:BHW:1} \\
\hspace{-0.75in}
P_{k}^{(N)} &=& \frac{\tilde{P}_{k}^{(N)}}{\sum_{k'=0}^{\np0}\,\tilde{P}_{k'}^{(N)}},  \quad
\tilde{P}_{k}^{(N)} =  (1-z)^N\, \frac{z^{k}}{(1-z^{\np0-k+1})^{N-1}}\,
\left(
       \begin{array}{c}
          k + N-1 \\
          k
        \end{array}
      \right).  \label{Pksi:corrected:BHW:2}
\eea  
In \Eq{Pksi:corrected:BHW:1} $\ket{\Phi^{(N)}_{k}}\equiv \ket{\Phi^{'(N)}_{k}}$ of \Eq{Phi:N:approx:BHW}, and we have simply removed the prime to indicate its association with the finer approximated probabilities in \Eq{Pksi:corrected:BHW:2}.
\Eq{Pksi:corrected:BHW} was  one of the  primary analytical results of this second model \cite{Alsing:2016}.
\begin{figure}[h]
\begin{center}
\begin{tabular}{cc}
\includegraphics[width=3.25in,height=2.0in]{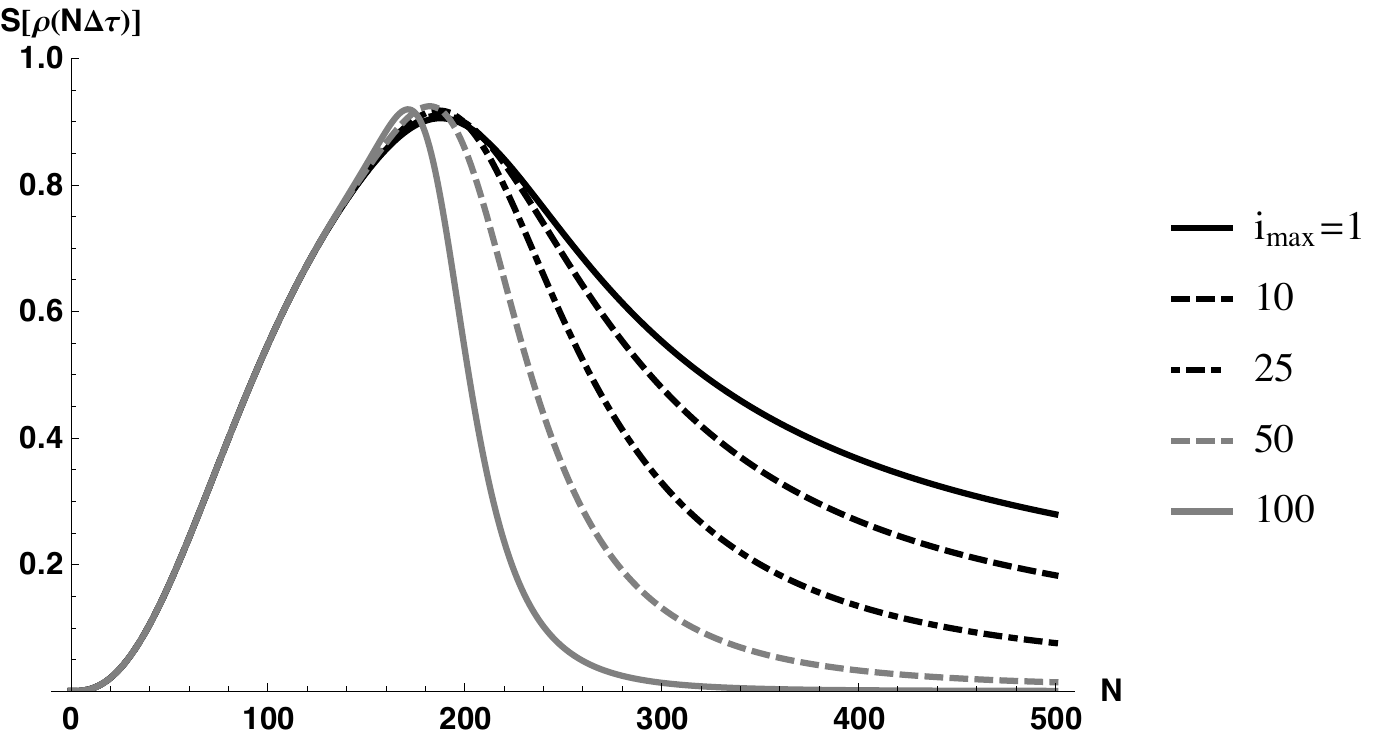}  &
\includegraphics[width=3.25in,height=2.0in]{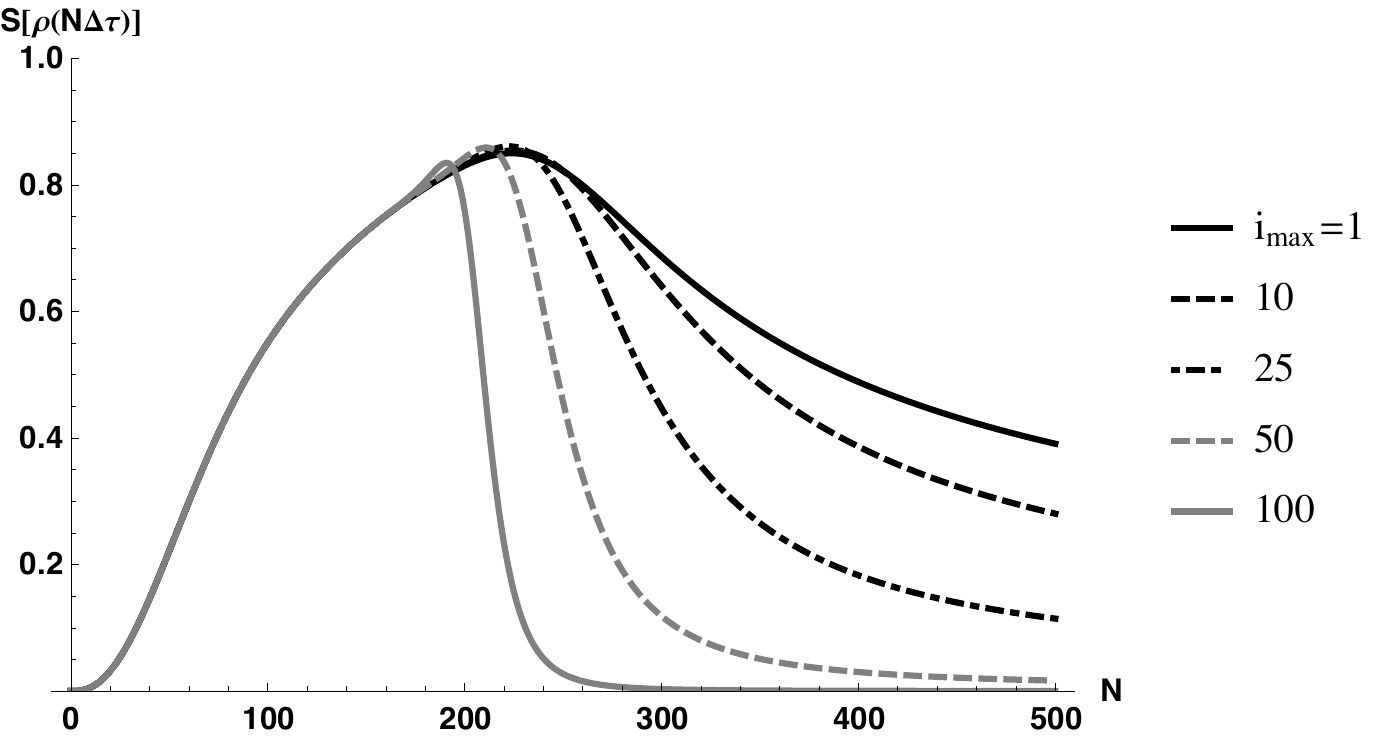}
\end{tabular}
\caption{Entropies with the extra term
$(1-z^{\np0-k+1})^{\textrm{min}(N-1,i_{max}-1)}$ in \Eq{Pksi:corrected:BHW:2} with various values of $i_{max}$ for
(left) $\np0=10$, (right) $\np0=25$.
Entropies are computed with $\log_{\np0+1}$ for the purpose of comparison.
}\label{fig:Pksi:imax:BHW}
\end{center}
\end{figure}

The effect of the extra factor $(1-z^{\np0-k+1})^{N-1}$ in $P_{k}^{(N)}$ can be seen by replacing it with
$(1-z^{\np0-k+1})^{min[N-1,i_{max}-1]}$ and varying the value of $i_{max} \le N$.
The value of $i_{max}$ sets how many terms $(1-z^{\np0-j_{i-1}+1})^{-1/2}$ in the $N-1$ nested sums in \Eq{Psi:j:jN:first:BHW} that we do not approximate as unity.
This is shown in \Fig{fig:Pksi:imax:BHW} for the cases of $\np0=10$ (left) and $\np0=25$ (right). These figures show how the additional factors
of  $(1-z^{\np0-k+1})$ in $P_{k}^{(N)}$ brings down the tail of entropy distribution $S$ to zero for longtimes, while leaving the short time (small $N$) portion of $S$ essentially unaltered.

\begin{figure}[h]
\begin{center}
\begin{tabular}{ccc}
%
\includegraphics[width=3.25in,height=2.0in]{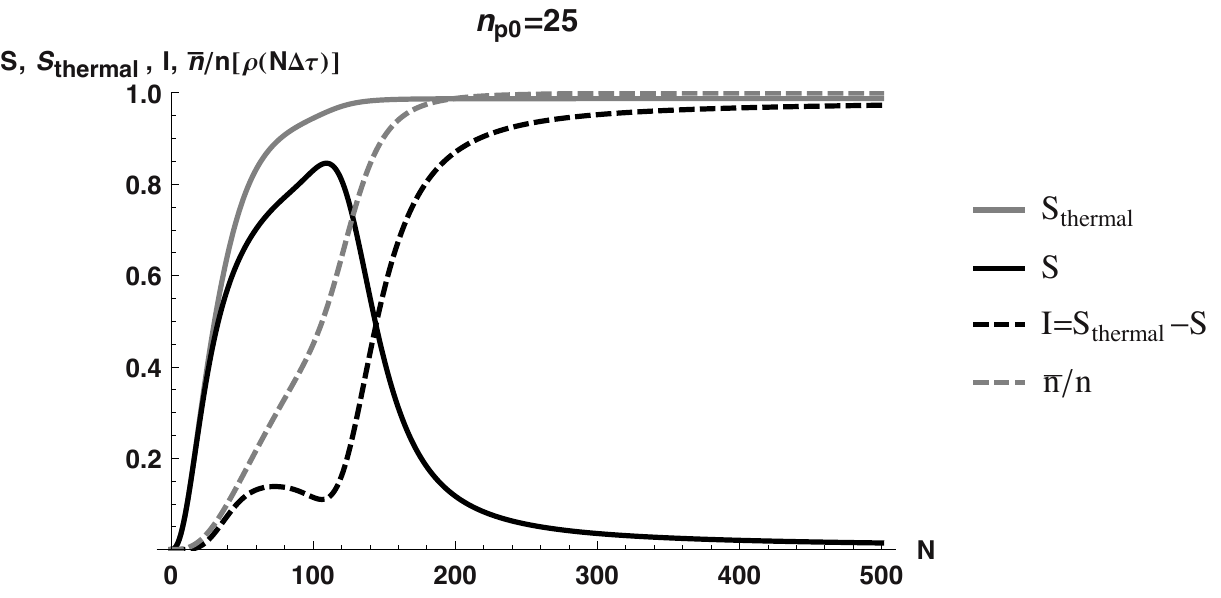} &
 {}& 
\includegraphics[width=3.25in,height=2.0in]{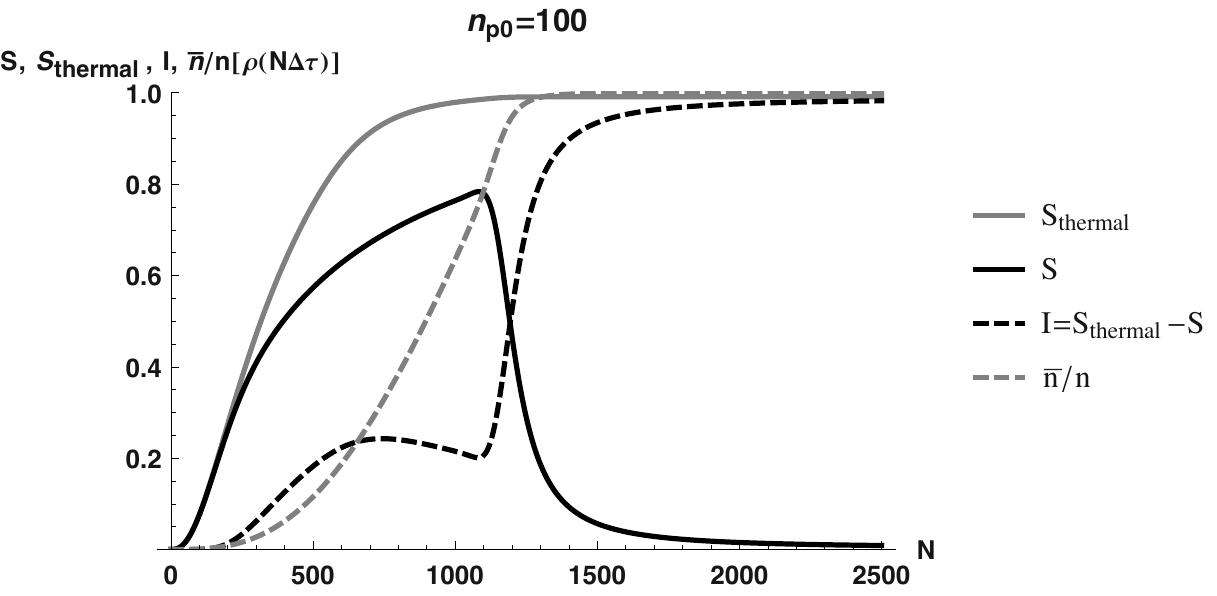} 
\end{tabular}
\caption{Plots of entropy $S$ (black, solid), effective $S_{thermal}$ (gray, solid), Page Information $I$ (black, dashed) and the fraction of emitted Hawking particles in signal/idler modes $\bar{n}_{s, i}/\np0$ vs time $N$
for 
(left) $\np0=25$, and
(right) $\np0=100$,
 with $z=0.1$, ($N_{z_{max}}=10^4,\, i_{max}=50$),
using the probabilities $P_k^{(N)}$ in \Eq{Pksi:corrected:BHW:2}.
Entropies are computed with $\log_{\np0+1}$ for the purpose of comparison.
}\label{fig:Allcurves:n25:n100:BHW}
\end{center}
\end{figure}
In \Fig{fig:Allcurves:n25:n100:BHW} we show plots of the entropy $S$ (black, solid), effective $S_{thermal}$ (gray, solid), Page Information $I=S_{thermal}-S$ (black, dashed) and the fractional number of the emitted Hawking particles in signal/idler modes $\bar{n}_{s, i}/\np0$ vs time $N$
using the probabilities $P_k^{(N)}$ in \Eq{Pksi:corrected:BHW:2} for 
(left) $\np0=25$ and  (right) $\np0=100$.
Entropies are computed with $\log_{\np0+1}$ so that all graphs have maximum value of unity, for comparison.
Both curves show that for early times (small $N$) $S\approx S_{thermal}$ so that the Page information $I$ is flat with, with very small slope. As time progresses, $I$ begins to grow, as $\bar{n}_{s, i}$ rapidly increases, and the BH begins to evaporate. For $\bar{n}_{s, i}/\np0 > 1/2$ there are less particles in the BH `pump' mode than have been emitted into all the Hawking radiation signal/idler modes and $S$ begins to decrease.
In \Fig{fig:Allcurves:n25:n100:BHW}(right)  for $\np0=100$ shows the initial flatness of the Page information $I$ is more pronounced.

Note that a brute force summation of all the terms in \Eq{Psi:j:jN:first:BHW} would involve the addition of
on the order of
$
\small{
\left(
       \begin{array}{c}
          k + N-1 \\
          k
        \end{array}
      \right)
}
$
summands, which equates to $10^{42}$ and $10^{182}$ terms for $k=\np0=25,\,N=500$ and $k=\np0=100,\,N=2500$ for
\Fig{fig:Allcurves:n25:n100:BHW}(left) and \Fig{fig:Allcurves:n25:n100:BHW}(right) respectively, which is impractical.
While most of the summands would be negligibly small and warrant approximating to zero,
a reasonable estimate of only $k=10$ nonzero terms per sum would still
lead to the prohibitive total number of nonzero summands of
$10^{20}$ and $10^{27}$ for $N=500$ and $N=2500$, respectively.
Hence, the necessity for the analytic approximations to the probabilities given
by \Eq{Psi:Normalized:BHW:2} and \Eq{Pksi:corrected:BHW:2} \cite{Alsing:2016}.

While this second model exhibits a Page curve, it has the same ``remnant issue" as the first model, namely only half the initial mass of the BH is emitted into the Hawking radiation at the end of the evaporation process. 
While the pump/BH reaches zero mass, the other have of the mass is contained in the idlers, trapped behind the horizon (assuming again for simplicity that $\om_s=\om_i=\thalf\,\om_p$), even though it is now distributed over a greater collection of different signal/idler modes. 
One of the main goals of the third model, considered in the next section, is to rectify this issue, while still preserving the Page curve feature of the second One Shot model.

\section{The Black Hole Waterfall Model}\label{sec:BHWModel}
We next consider the third model, which we descriptively term the Black Hole Waterfall model (BHW), which is the main focus of this present work. The BHW model consists of a generalization of the first model, considered in \Sec{subsec:review:Alsing:2015}, to which the One Shot apparatus of the second model, considered in \Sec{subsec:OneShot:2016}, is applied, plus the additional feature whereby the interior idler/Hawking partner particles can act as subsequent SPDC sources themselves.

It is notationally a bit cumbersome to write down and interpret at first glance in complete generality, so we will first build it up in pieces, and then present the main analytic results used to create subsequent entropy plots. A full derivation is provided in  \ref{app:BHW:derivation}, and mimics the derivation of the second model in \Sec{subsec:OneShot:2016}.

\subsection{The Basic Black Hole Waterfall Mechanism}\label{subsec:BHW:Without:OneShot}
As discussed at the end of \Sec{subsec:review:Alsing:2015}, the main feature of the BHW model is that we allow generated idler (the Hawking partner particle behind the BH horizon) to be able to act as its own pump, thus generating subsequent signal/idler pair of half the energy of the idler pump (when we assume degenerate SPDC, for ease of discussion). Thus, going back to the first model in \Sec{subsec:review:Alsing:2015}
\Eq{psit:BHW} evolves in sequence of cascading idler pumps labeled by the index $D$ (for \tit{depth}) as 
\smallskip

\bea{psit:D:example:BHW}
\hspace{-1.25in}
 \ket{\psi_{D=1}(\tau)} &=& 
\sum_{n_1=0}^{\np0} \, 
c_{n_1}(\tau)\, \Big[\ket{\np0-n_1}_p\ket{n_1}_{i_1}\ket{n_1}_{s_1}\Big], 
\label{psit:D:example:BHW:D:1} \\
\hspace{-1.25in}
\ket{\psi_{D=2}(\tau)} &=& 
\sum_{n_1=0}^{\np0}   \sum_{n_2=0}^{n_1} \, 
c_{n_1 n_2}(\tau)\, \ket{\np0-n_1}_p
\Big[\ket{n_1-n_2}_{i_1}\ket{n_2}_{i_2}\ket{n_2}_{s_2}\Big]
\ket{n_1}_{s_1}, 
 \label{psit:D:example:BHW:D:2} \\
 \hspace{-1.25in}
 \ket{\psi_{D=3}(\tau)} &=& 
\sum_{n_1=0}^{\np0}    \sum_{n_2=0}^{n_1}    \sum_{n_3=0}^{n_2} 
c_{n_1 n_2 n_3}(\tau)\, \ket{\np0-n_1}_p\ket{n_1-n_2}_{i_1}
\Big[\ket{n_2-n_3}_{i_2}\ket{n_3}_{i_3}\ket{n_3}_{s_3} \Big]
\ket{n_2}_{s_2}\ket{n_1}_{s_1}.\qquad
 \label{psit:D:example:BHW:D:3} 
\eea

\begin{figure}[h]
\begin{center}
\includegraphics[width=4in,height=4.0in]{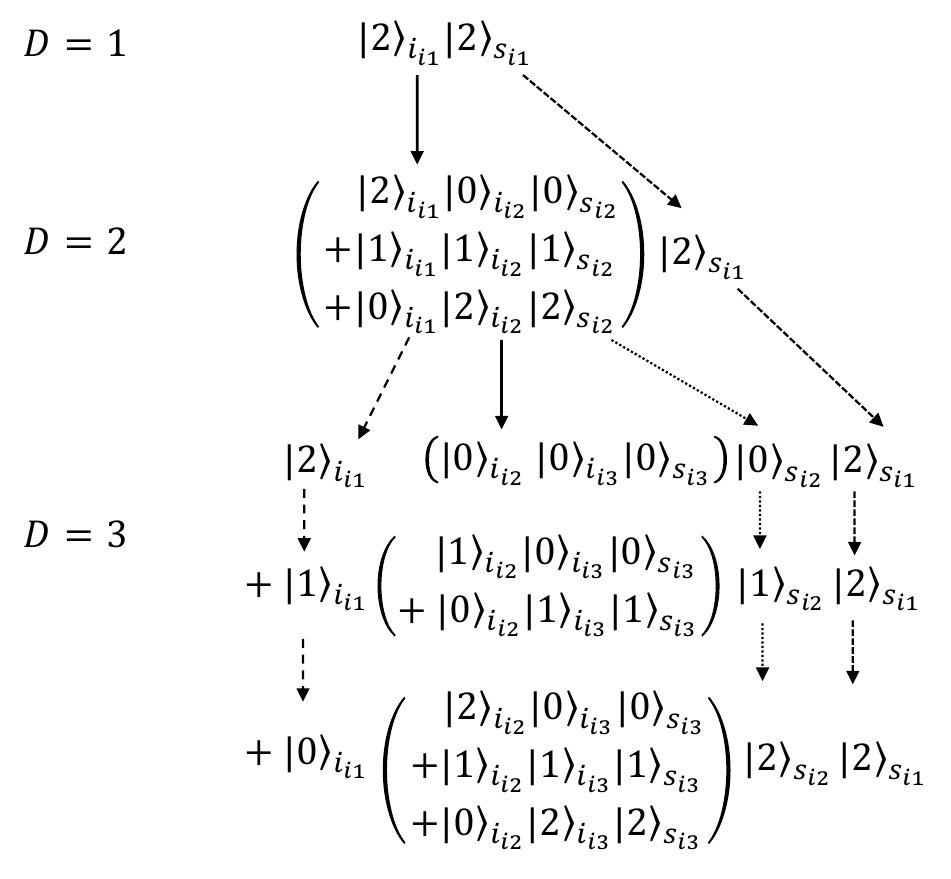} 
\caption{A visual representation of the subsequent signal/idler states generated by the idler pumps 
from a state of a signal idler pair $\ket{2}_{i_{i1}}\ket{2}_{s_{i1}}$ 
(created by the pump from the component $\ket{\np0-2}_p$ appearing at time step $i\in\{1,2,\ldots,N\}$ for $D=(1,2,3)$.
The solid black arrow indicates how idler pump $i_{i1}$ at $D=1$ is the source of signal/idler pairs $(i_{i2},s_{i2})$ at $D=2$.
In turn, the idler $i_{i2}$ is the pump source of signal/idler pairs $(i_{i3},s_{i3})$ at $D=3$, etc\ldots.
}\label{fig:Deq123:cartoon:BHW}
\end{center}
\end{figure}

Note that in \Eq{psit:D:example:BHW:D:3} we have $n_1\ge n_2 \ge n_3$. 
The indices $n_1\ge n_2\ge n_3$ and idler/signal states in appearing in  
\Eq{psit:D:example:BHW:D:3} for $D=3$ for $k_1=(0,1,2)$ (i.e up to 2 signal/idler pairs emitted by the pump/BH; 
$\ket{\np0-k_1}_p$) are shown in
\Tbl{tbl:n1n2n3:signal:idler:states:BHW}.
\begin{table}[h!]
\begin{center}
\begin{tabular}{|c|c|c|} \hline
\multicolumn{3}{|c|}{Indices and states in appearing in  \Eq{psit:D:example:BHW:D:3} for $D=3,\; k_1=(0,1,2)$} \\ \hline
\multicolumn{1}{|c|}{$n_1\ge n_2\ge n_3$} & 
\multicolumn{1}{|c|}{$\ket{n_1-n_2}_{i_1}\ket{n_2-n_3}_{i_2}\ket{n_3}_{i_3}$ } & 
\multicolumn{1}{|c|}{$\ket{n_3}_{s_3}\ket{n_2}_{s_2}\ket{n_1}_{s_1}$} \\ \hline\hline
$0,0,0$ & $\ket{0}_{i_1}\ket{0}_{i_2}\ket{0}_{i_3}$ & $\ket{0}_{s_3}\ket{0}_{s_2}\ket{0}_{s_1}$ \\ \hline
$1,0,0$ & $\ket{1}_{i_1}\ket{0}_{i_2}\ket{0}_{i_3}$ & $\ket{0}_{s_3}\ket{0}_{s_2}\ket{1}_{s_1}$ \\
$1,1,0$ & $\ket{0}_{i_1}\ket{1}_{i_2}\ket{0}_{i_3}$ & $\ket{0}_{s_3}\ket{1}_{s_2}\ket{1}_{s_1}$ \\
$1,1,1$ & $\ket{0}_{i_1}\ket{0}_{i_2}\ket{1}_{i_3}$ & $\ket{1}_{s_3}\ket{1}_{s_2}\ket{1}_{s_1}$ \\ \hline
$2,0,0$ & $\ket{2}_{i_1}\ket{0}_{i_2}\ket{0}_{i_3}$ & $\ket{0}_{s_3}\ket{0}_{s_2}\ket{2}_{s_1}$ \\
$2,1,0$ & $\ket{1}_{i_1}\ket{1}_{i_2}\ket{0}_{i_3}$ & $\ket{0}_{s_3}\ket{1}_{s_2}\ket{2}_{s_1}$ \\
$2,1,1$ & $\ket{1}_{i_1}\ket{0}_{i_2}\ket{1}_{i_3}$ & $\ket{1}_{s_3}\ket{1}_{s_2}\ket{2}_{s_1}$ \\
$2,2,0$ & $\ket{0}_{i_1}\ket{2}_{i_2}\ket{0}_{i_3}$ & $\ket{0}_{s_3}\ket{2}_{s_2}\ket{2}_{s_1}$ \\
$2,2,1$ & $\ket{0}_{i_1}\ket{1}_{i_2}\ket{1}_{i_3}$ & $\ket{1}_{s_3}\ket{2}_{s_2}\ket{2}_{s_1}$ \\
$2,2,2$ & $\ket{0}_{i_1}\ket{0}_{i_2}\ket{2}_{i_3}$ & $\ket{2}_{s_3}\ket{2}_{s_2}\ket{2}_{s_1}$ \\
\hline
\end{tabular}
\end{center}
\caption{The indices $n_1\ge n_2\ge n_3$ and idler/signal states in appearing in  
\Eq{psit:D:example:BHW:D:3} for $D=3$ with $k_1=(0,1,2)$ (i.e. up to two Hawking pairs emitted by the BH; $\ket{\np0-k_1}_p$).
}\label{tbl:n1n2n3:signal:idler:states:BHW}
\end{table}
A visual representation of the generation of the subsequent signal/idler pairs from a pair $\ket{2}_{i_1}\ket{2}_{s_1}$ created from the BH component term $\ket{\np0-2}_p$ is shown in \Fig{fig:Deq123:cartoon:BHW}.

The large square brackets in \Eq{psit:D:example:BHW:D:1}--\Eq{psit:D:example:BHW:D:3} 
are placed to help visualize how idler pump $\ket{n_{d-1}}_{i_{d-1}}$  
at each intermediate depth $d\in[1,D]$ of frequency  $\om_{i_{d-1}}$ 
(defining $i_0\leftrightarrow p$) 
generates its own subsequent signal/idler pair
 $\ket{n_{d}}_{i_{d}}\ket{n_{d}}_{s_{d}}$ 
 of frequency  $\om_{i_d}$ and $\om_{s_{d}}$ 
 by emitting $n_{d}$ excitations, i.e.
  $\ket{n_{d-1}}_{i_{d-1}}\to \ket{n_{d-1}-n_{d}}_{i_{d-1}}$, and conserving energy in the form 
  $\om_{i_{d-1}} = \om_{i_{d}} +\om_{s_{d}}$.
  Thus, the system evolves under the Hamiltonian
  \be{Hamiltonian:BHW}
  H = i\, \Big(
  \sum_{d=1}^{D} a_{i_{d-1}}\,a^\dag_{i_d}\,a^\dag_{s_d} - a^\dag_{i_{d-1}}\,a_{i_d}\,a_{s_d}
  \Big), \qquad a_{i_0}\equiv a_p.
  \ee
   It is then easy to see from the above pattern of occupation numbers that the total energy $\np0\,\om_p$ is conserved at any total depth $D$ since
%

   \bea{energy:conservation:D:BHW}
  \hspace{-0.5in}
 &{}& \sum_{d=1}^{D} (n_{d-1} - n_{d})\,\om_{i_{d-1}} + n_{i_D}\, \om_{i_D} + 
 \sum_{d=1}^{D} n_{d}\, \om_{s_d}, \no 
 %
  \hspace{-0.5in}
 &=& n_{i_0} \,\om_{i_0} + \sum_{d=1}^{D} n_d\,(-\om_{i_{d-1}}+ \om_{i_d} + \om_{s_d}) \equiv \np0\,\om_p, \quad
 \trm{using}\; \om_{i_{d-1}} = \om_{i_{d}} +\om_{s_{d}},\qquad
  \eea
 where in \Eq{energy:conservation:D:BHW} we have defined $n_0\equiv \np0$ and $\om_{i_0}\equiv \om_p$.
  
Henceforth, without loss of generality and for ease of discussion, it's easiest to consider the 
most symmetric degenerate SPDC case where  $\om_{i_{d}}=\om_{s_{d}} = \thalf \om_{i_{d-1}}$. 
The state $\ket{\psi_{D}(\tau)}$ admits an initial state at $\tau=0$ with all $n_d = 0$, 
$d\in[1,D]$, namely
$\ket{\psi_{D}(0)}= \ket{\np0}_p\,\prod_{d=1}^D\ket{0}_{i_d}\ket{0}_{s_d}$, with the BH having energy (mass) $\np0\om_p$.
More importantly, the state $\ket{\psi_{D}(\tau)}$ admits the desired evaporated state with all $n_d = \np0$, namely
(see, e.g. \Eq{psit:D:example:BHW:D:3})
\be{psi:evap:state:BHW}
\ket{\psi_{D}^{(evap)}}= \ket{0}_p\otimes\prod_{d=1}^{D-1}\ket{0}_{i_d}\otimes \ket{\np0}_{i_D}\otimes\prod_{d=1}^D\ket{\np0}_{s_d},
\ee
with 
(i) the pump/BH in its vacuum state, 
(ii) all the interior idlers/Hawking partner particles inside the horizon being in their vacuum date \tit{except} the $D$-th mode at frequency $\om_p/2^D$, and
(iii)  \tit{all} the signal modes having occupation number $\np0$. Thus, energy distributed all the external signal modes/Hawking radiation is $\sum_{d=1}^D (\thalf)^d\, \np0\om_p = \left(1 - (\thalf)^D\right)  \np0\om_p$, while behind the horizon the pump/BH has zero mass, and we are left with a vanishingly small energy  $(\thalf)^D \np0\om_p$ in internal idler/Hawking partner particles in mode $i_D$. In the limit $D\to\infty$ these energies asymptote to the entire initial mass $ \np0\om_p$ of the BH appearing the Hawking radiation, and  zero mass left in the interior Hawking partner particles. This  ``waterfall" of cascading idler pumps is qualitatively reminiscent of ``ER=EPR" and Replica Wormhole/Island mechanism discussed in the introduction in the sense that both function to move interior Hawking particle particles behind the horizon unitarily to the external Hawking radiation as the evaporation evolves. However, the BHW mechanism need not invoke wormholes as a physical process. The BHW mechanism's possible relationship to the Island effect is unclear at this point, though remains an intriguing avenue for further investigation.
In the next section we apply the One Shot Trotterization of the second model \cite{Alsing:2016} to this BHW mechanism.

\begin{figure}[h]
\hspace{0.75in}
\includegraphics[width=6.0in,height=2.5in]{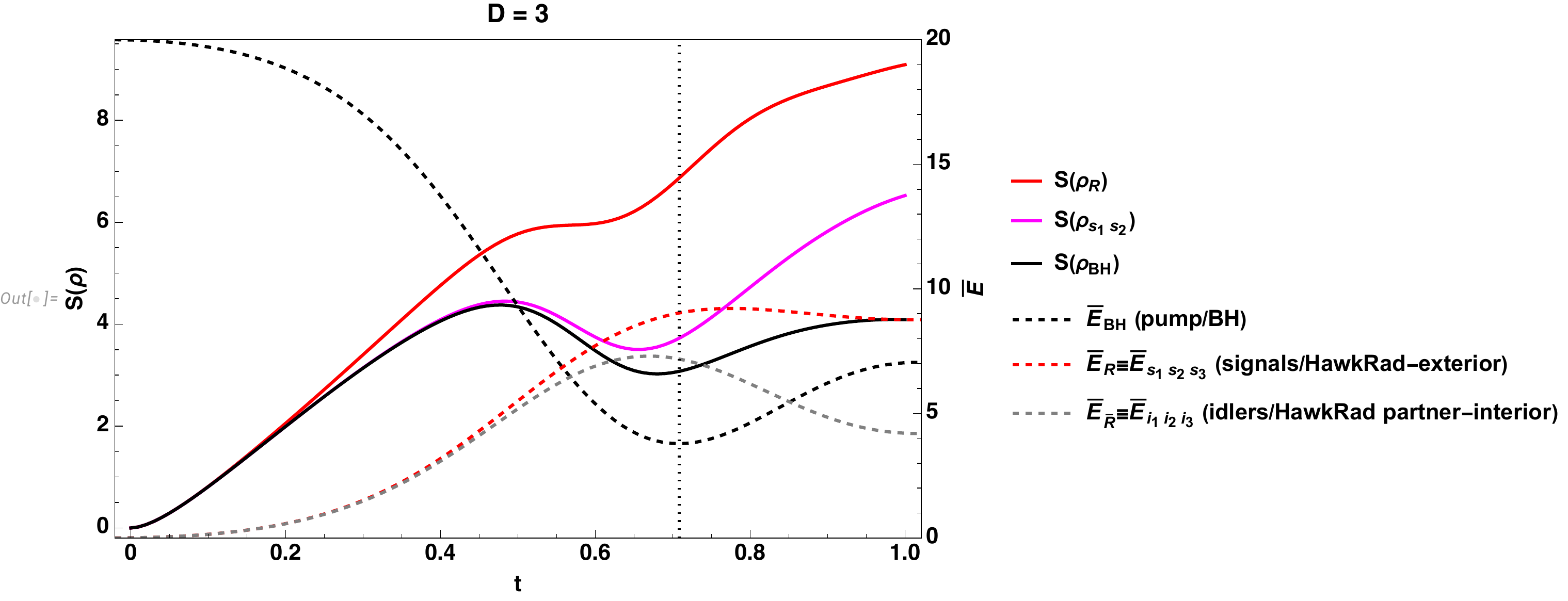} 
\caption{Direct numerical integration of \Eq{psit:D:example:BHW:D:3} for the case of $D=3, \np0=20$,  $z~=~0.1$ (without the One Shot mechanism). Entropy (solid) curves $S(\rho)$ are plot using the left ordinate axis, mean energy (dashed) curves $\bar{E}$ are plotted using the right ordinate axis. See the text for a description. 
}\label{fig_entropies_Ebars_Dimeq3np0eq20tmaxeq1dteq0p1_15Dec2024:BHW}
\end{figure}

Before moving on, we first note that 
we can numerically integrate the Schr\"{o}dinger equation for a given $D$, but the number of amplitudes grows as $(\np0)^D$.
In \Fig{fig_entropies_Ebars_Dimeq3np0eq20tmaxeq1dteq0p1_15Dec2024:BHW} we perform the direct numerical integration of \Eq{psit:D:example:BHW:D:3} for the case of $D=3$. Because the number of amplitudes scales as 
$
{
\tiny
\left(
           \begin{array}{c}
             D+\np0-1 \\
             D
           \end{array}
\right)
},
$
we lower $\np0=20$ to keep the integration time and memory to a reasonable size. Again, the entropy curves have the same qualitative shapes for larger values of  $\np0$, but the scale of the time axis grows as $\sqrt{\np0}\,$.

Because the pure state density matrix is now a composite system of the pump/BH and all the signal (exterior Hawking radiation) and idlers (interior Hawking particles),
$\rho_{p,i_1,i_2,i_3,s_1,s_2,s_3}\equiv \rho_{p,I,S}$ we now have 
$S(\rho_{p}) \equiv  S(\rho_{I,S}) \ne S(\rho_{S})$. In fact it is easy to see from \Eq{psit:D:example:BHW:D:3} that 
\bea{rho:p:S:D3:tmax1:BHW}
\fl \rho_{p}=\rho_{BH} &=& \sum_{n_1=0}^{\np0}   
 \left( 
 \sum_{n_2=0}^{n_1}    \sum_{n_3=0}^{n_2}
 |c_{n_1 n_2 n_3}|^2
 \right)\, 
 \ket{\np0-n_1}_p \bra{\np0-n_1}_p, \label{rho:p:S:D3:tmax1:BHW:p}\\
\fl  \rho_{S}=\rho_{s_1,s_2,s_3} &=&   
 \sum_{n_1=0}^{\np0} \sum_{n_2=0}^{n_1}    \sum_{n_3=0}^{n_2}
  |c_{n_1 n_2 n_3}|^2\, 
  \ket{n_1,n_2,n_3}_{s_1,s_2,s_3}  \bra{n_1,n_2,n_3}, \label{rho:p:S:D3:tmax1:BHW:S}\\
\fl &{}& 
\hspace{-0.55in}
\rho_{s_1,s_2} =   
 \sum_{n_1=0}^{\np0} \sum_{n_2=0}^{n_1}    
 \left(
 \sum_{n_3=0}^{n_2}
  |c_{n_1 n_2 n_3}|^2
  \right)\,
  \ket{n_1,n_2}_{s_1,s_2}  \bra{n_1,n2}, \label{rho:p:S:D3:tmax1:BHW:ps1s2}
 \eea
which are plotted in \Fig{fig_entropies_Ebars_Dimeq3np0eq20tmaxeq1dteq0p1_15Dec2024:BHW}
in $S(\rho_{BH}) $ (solid black curve), $S(\rho_{s_1,s_2})$ (solid magenta curve), and 
$S(\rho_{S} )$ (solid red curve).
The dashed curves show the evolution of 
$\bar{E}_{BH}(\tau)$ (dashed black) the BH mass/energy, 
$\bar{E}_{R}(\tau)$ (dashed red) the exterior Hawking radiation (signals), and $\bar{E}_{\bar{R}}(\tau)$ (dashed gray) the interior Hawking partner particles (idlers). While the same basic probabilities 
$|c_{n_1 n_2 n_3}|^2$ are involved in all three computed density matrices, each involves a different coarse graining of the basic probabilities, and hence yield different entropy curves.

Again, the model should not really be applied past the point at which 
$d\bar{n}_{BH}(\tau)/d\tau=0$, shown as the vertical dotted black line in \Fig{fig_entropies_Ebars_Dimeq3np0eq20tmaxeq1dteq0p1_15Dec2024:BHW}. As in first model in \Sec{subsec:review:Alsing:2015}, the entropy curve for the BH begins to turn over at this point. However, beyond this point, oscillations between the Hawking radiation and the BH begin to set in, which will be removed by applying the One Shot mechanism in the next section, to allow the Hawking radiation to escape to infinity (by effectively not interacting substantially with the pump/BH, once created).

\subsection{The One Shot Black Hole Waterfall Model}\label{subsec:BHW:With:OneShot}
We now apply the One Shot mechanism to the BH pump combined with the (depth) $D$ additional idler ``pumps." 
A more detailed derivation is given in the Appendix; here we state the primary final analytical results.
Note that we implicitly assume here that the waterfall process occurs rapidly, all at once, at each time step, i.e. on a timescale faster the time between pump/BH emission events in the  One Shot (second) model.

Each pump (source of SPDC generation) is associate with a probability that is a generalization of 
$P_{k}^{(N)}$ in \Eq{Pksi:corrected:BHW:2}, and are given by
\bea{P:kd:kdm1:BHW}
\fl P_{k_{d}}^{(k_{d-1})} &=& \frac{\tilde{P}_{k_{d}}^{(k_{d-1})}}{\sum_{k'_d=0}^{k_{d-1}}\, \tilde{P}_{k'_{d}}^{(k_{d-1})} },  \quad
\tilde{P}_{k_d}^{(k_{d-1})} =  (1-z)^N\, \frac{z^{k_d}}{(1-z^{k_{d-1}-k_{d}+1})^{N-d}}\,
\left(
       \begin{array}{c}
          k_d + N-1 \\
          k_d
        \end{array}
      \right). \qquad
\eea      
Essentially,  after ``time" $N$ (which counts the BH emission events)
and at depth $d\in[1,D]$, we have the substitutions $k\to k_d$, $\np0-k+1\to k_{d-1} - k_{d}+1$ 
in the probabilities \Eq{Pksi:corrected:BHW:2}, $P_{k}^{(N)}\to P_{k_{d}}^{(k_{d-1})}$ 
(defining $k_0\equiv \np0$).
The wavefunction after $N$ pump iterations, with a depth $D$ of subsequent idler pump emission events is given by
\bea{psi:N:D:BHW}
\hspace{-1.0in}
 \ket{\psi^{(N)}_D} \approx 
\sum_{k_1=0}^{\np0}\,\sqrt{P_{k_1}^{(\np0)}}\,
\sum_{k_2=0}^{k_1}\,\sqrt{P_{k_2}^{(k_1)}}\,
\sum_{k_3=0}^{k_2}\,\sqrt{P_{k_3}^{(k_2)}}\,
\cdots
\sum_{k_D=0}^{k_{D-1}}\,\sqrt{P_{k_D}^{(k_{D-1})}}\,
\ket{\np0-k_1}_p\otimes\ket{\Phi^{(N)}_{k_1,k_2,\ldots,k_D}}. \quad\quad
\eea
Here the $k_d$  with $d\in[1,D]$ are the collective variables representing the total number of particles emitted by idler pump $i_{d-1}$ (with idler pump $i_0$ defined to be the BH) into subsequent downstream signal/idler Hawking pairs. Thus, $k_1\in[0,\np0]$ is the total number of Hawking pairs emitted by the BH into \tit{all} the interior/exterior Hawking particles. The upper bound on the summations over $k_d$ for a given idler pump $i_{d-1}$, is given by  the number of particles $k_{d-1}$ that
were initially in the idler pump source $i_{d-1}$.

In the  approximation used for \Eq{psi:N:D:BHW}, 
the normalized orthogonal states $\ket{\Phi^{(N)}_{k_1,k_2,\ldots,k_D}}$ are a generalization of the states (with $j_N\to k_1$)
$\ket{\Phi^{(N)}_{k_1}}$  appearing in \Eq{Phi:N:approx:BHW} (the latter of which has \Eq{Phi:N4:jN2:BHW}  is a particular example).
In fact, the state  $\ket{\Phi^{(N)}_{k_1}}$ in  \Eq{Phi:N:approx:BHW}  is simply the state if we had only a pump-depth of $D=1$, which is just the second model, where only the BH (and no idlers) acts as a SPDC source. For $D>1$, each idler $i_{i,{d-1}}$ mode (where $i\in[1,N]$ and $d\in[1,D]$) generates further signal/idler pairs $(i_{i,d}, s_{i,d})$ as indicated in \Tbl{tbl:n1n2n3:signal:idler:states:BHW}, and illustrated in \Fig{fig:Deq123:cartoon:BHW}. The index $k_d$
indicate the total number of signal/idler pairs generated by idler-pump $i_{d-1}$ (e.g, $i_0$ is defined to be the BH pump).

For example, the  component $\ket{2}_3 \equiv \ket{2}_{i_{31}}\ket{2}_{s_{31}}$ of the state $\ket{0,0,2,0}_{1,2,3,4}$ appearing in the $D=1$ example state $\ket{\Phi^{(N=4)}_{k_1}}$ in  \Eq{Phi:N4:jN2:BHW} would be expanded as in the last row of  \Fig{fig:Deq123:cartoon:BHW}  for $D=3$ (after substituting $i\to 3$ there). Each component of all of the ten 4-kets appearing in the state $\ket{\Phi^{(N=4)}_{k_1}}$ in  \Eq{Phi:N4:jN2:BHW}  would be expanded in a similar fashion for a pump-depth of $D=3$, contributing to the collective excitation state $\ket{\Phi^{(N=4)}_{k_1,k_2,k_3}}$ in \Eq{psi:N:D:BHW}. 

Note that from \Eq{psi:N:D:BHW}  
$\rho_{BH}\equiv\rho_p = \sum_{k_1=0}^{\np0} P_{k_1}^{(\np0)}\ket{\np0-k_1}_p\bra{\np0-k_1}$ since the 
$\ket{\Phi^{(N)}_{k_1,k_2,\ldots,k_D}}$ are orthonormal, and all the sums over the $D-1$ probabilities 
$P_{k_d}^{(k_{d-1})}$ sum to unity (by construction), i.e $\sum_{k_d}^{k_{d-1}} P_{k_d}^{(k_{d-1})} = 1$. Thus, the density matrix for the BH is independent of $D$. This makes sense since 
$k_1\in\{0,1,\ldots,\np0\}$ represents the total number of $(i_1,s_1)$ Hawking pairs 
created by the pump/BH, which subsequently may or may not act like downstream pump  sources depending on whether or not we invoke the waterfall mechanism (i.e. idlers acting as SPDC pump sources). On the other hand, the density matrix for the Hawking radiation (all the signal modes) contains a nested sum over all the probabilities $P_{k_d}^{(k_{d-1})}$, given by 
\bea{rhoBH:rhoR:BHW}
\fl \rho_{BH} \equiv \rho_{p} &=& 
 \sum_{k_1=0}^{\np0}\, P_{k_1}^{(\np0)}\,
 \Big(
 \sum_{k_2=0}^{k_{1}}P_{k_2}^{(k_{1})}
 \cdots\sum_{k_D=0}^{k_{D-1}}\, P_{k_D}^{(k_{D-1})}
 \Big)
\ket{\np0-k_1}_p\bra{\np0-k_1}, \no
&=& \sum_{k_1=0}^{\np0}\, P_{k_1}^{(\np0)}\,\ket{\np0-k_1}_p\bra{\np0-k_1}, \qquad \label{rhoBH:rhoR:BHW:rhoBH}  \\
\fl \rho_R \equiv \rho_{s_1,\ldots,s_D} &=& 
 \sum_{k_1=0}^{\np0}\, \sum_{k_2=0}^{k_{1}}\cdots\sum_{k_D=0}^{k_{D-1}}\,
 P_{k_1}^{(\np0)}\,P_{k_2}^{(k_{1})}\cdots P_{k_D}^{(k_{D-1})}\,
 \Tr_I\left[ \ket{\Phi^{(N)}_{k_1,k_2,\ldots,k_D}}\bra{\Phi^{(N)}_{k_1,k_2,\ldots,k_D}}\right]. \qquad
\label{rhoBH:rhoR:BHW:rhoR} 
\eea
While these are both diagonal density matrices 
(see \Eq{rho:p:S:D3:tmax1:BHW:p} and \Eq{rho:p:S:D3:tmax1:BHW:S}), 
the probabilities involved are very different, and thus 
$S(\rho_{BH})\ne S(\rho_{R})$, even though the composite state is pure 
(however, purity of the composite state does imply that $S(\rho_{BH}) = S(\rho_{I,S})$, where
$S=({s_1,\ldots,s_D})$ and $I=({i_1,\ldots,i_D})$).
The BH $\bar{E}_{BH}$ and Hawking radiation $\bar{E}_{R}$ (all signals) energies, in units of 
$\om_p\to1$ are given by taking the expectation values of $\langle a^\dag_p\,a_p\rangle$ and
$\langle \sum_{d=1}^{D} a^\dag_{s_d}\,a_{s_d}/2^d\rangle$, respectively yielding
\bea{EbarBH:EbarR:BHW}
\hspace{-0.5in}
\bar{E}_{BH} &=& \sum_{k_1=0}^{\np0} P_{k_1}^{(\np0)}, \label{EbarBH:BHW} \\
\hspace{-0.5in}
 \bar{E}_{R} &=& 
 \sum_{k_1=0}^{\np0}\, \sum_{k_2=0}^{k_{1}}\cdots\sum_{k_D=0}^{k_{D-1}}\,
\left(
\frac{k_1}{2} + \frac{k_2}{2^2} +\cdots +\frac{k_D}{2^D}
\right)
P_{k_1}^{(\np0)}\,P_{k_2}^{(k_{1})}\cdots P_{k_D}^{(k_{D-1})}.  \label{EbarR:BHW}
\eea 

\subsection{Numerical results for the One Shot Black Hole Waterfall Model}\label{subsec:Numerics:BHW}
In this section we show numerical results for the 
BH entropy $S(\rho_{BH})$ computed from \Eq{rhoBH:rhoR:BHW:rhoBH}, and the 
Hawking radiation $S(\rho_{R})$ computed from \Eq{rhoBH:rhoR:BHW:rhoR} for various pump-depths $D$.
We also invoke the first law of (BH) thermodynamics and define the temperature of the BH as
\be{Temp:defn:BHW}
dE = T\,dS \quad \Rightarrow\quad T \equiv \frac{dE_{BH}}{dS_{BH}} = \frac{dE_{BH}/dx}{dS_{BH}/dx},
\ee
where $x$ is the continuous, interpolated ``time" formed from the discrete evolution time $N$ in the One Shot mechanism.

\subsubsection{$\mathbf{D=1:}$}\label{subsubsec:Deq1} 
\begin{figure}[h]
\begin{tabular}{ccc}
\hspace{-0.8in}
%
\includegraphics[width=4.0in,height=2.25in]{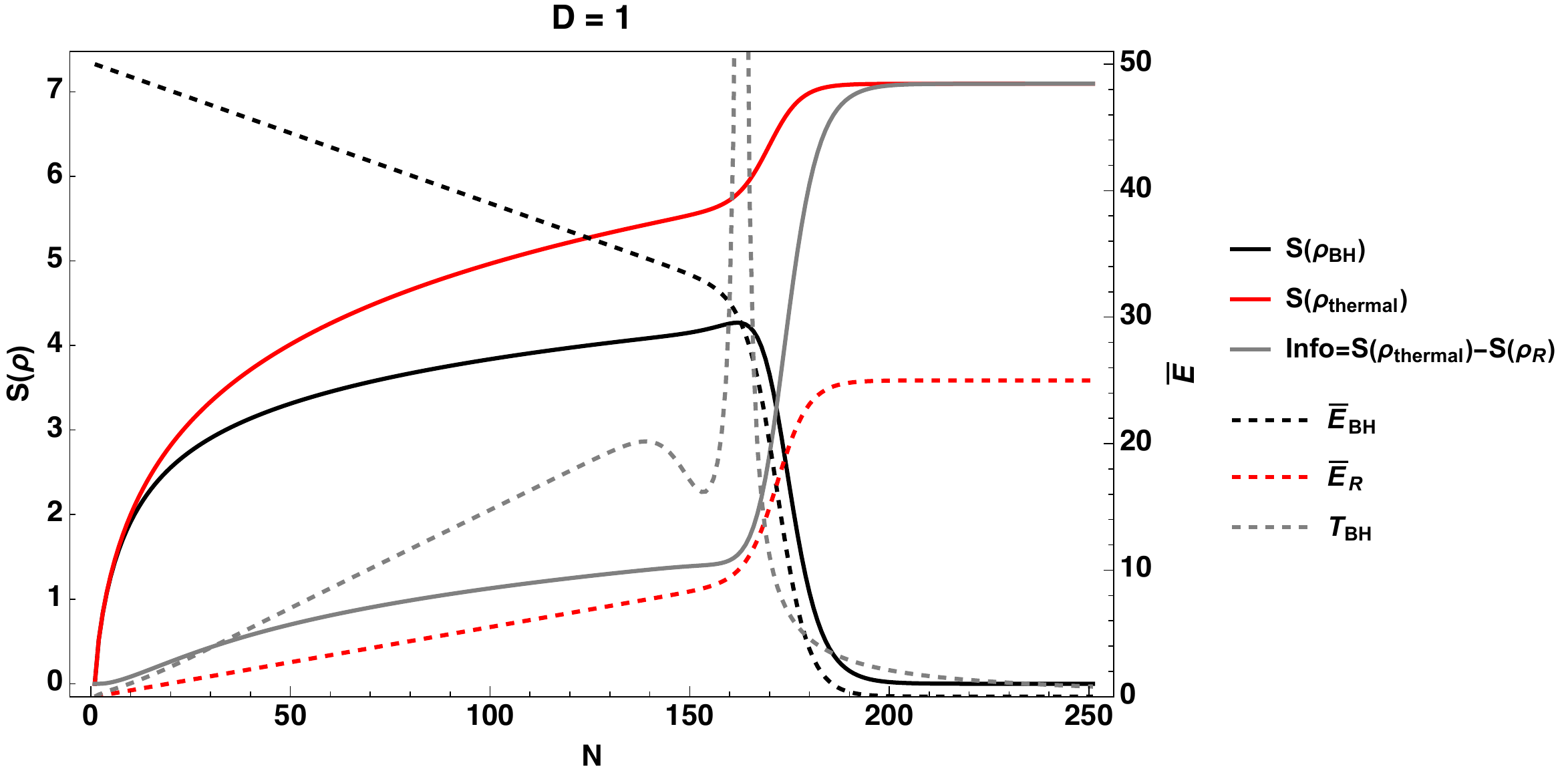} &
{}& 
\hspace{-0.35in}
\includegraphics[width=4.0in,height=2.25in]{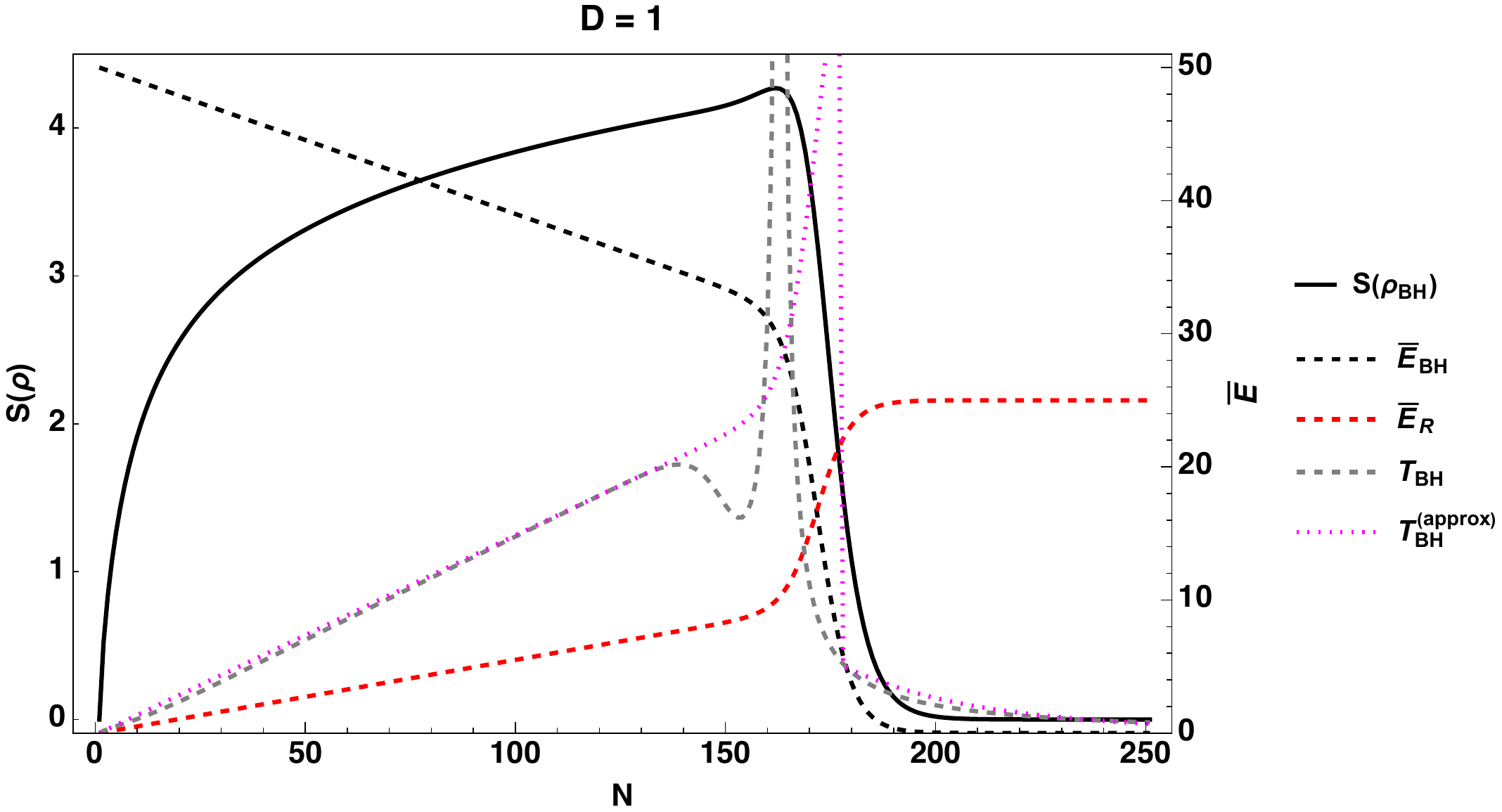}
%
\end{tabular}
\caption{(left)  Black hole entropy (black solid), effective thermal entropy (red solid) and Page Information (gray solid) for $D=1, \np0=50, z=0.1$.  Black hole energy (black dashed), Hawking radiation energy (red dashed), and black hole temperature (gray dashed)  $T_{BH} \equiv dE_{BH}/dS_{BH}$.
(right)~Black hole temperature from $T_{BH}$ (gray dashed), and analytic approximations
(magenta dotted) fitted for early ($N\in[1,139]$) and late ($N~\in~[178,251]$) times.
}\label{fig_Deq1np050Nmaxeq250_combine_plots_25Dec2024_TempBHvsEBH_26Dec2024:BHW}
\end{figure}

 In \Fig{fig_Deq1np050Nmaxeq250_combine_plots_25Dec2024_TempBHvsEBH_26Dec2024:BHW}(left) we plot the 
 entropy $S(\rho_{BH})$ (black solid), energy $\bar{E}_{BH}$ (black dashed) and temperature $T_{BH}$ (gray dashed) for the case $D=1$, $z=0.1$, with $\np0=~50~=~\bar{E}_{BH}(0)$ (setting $\om_p\to1$), along with the energy of the Hawking radiation $\bar{E}_{R}$ (red dashed), the effective thermal entropy 
 $S(\rho_{thermal})$ (red solid) constructed from the probabilities $p_n^{(thermal)} = (\nbarp)^n/(\nbarp+1)^{n+1}$ from the mean number $\nbarp$ of particles in the pump/BH, and the Page Information \cite{Nation_Blencowe:2010,Alsing:2016} defined
 as $I(N) = S(\rho_{thermal})-S(\rho_{BH})$. Note that the energy $\bar{E}_{R}$ of the Hawking radiation (red dashed curve) reaches only half the energy $\bar{E}_{BH}$  of the BH for $D=1$, since the other half is contained within the idlers/Hawking partner particles behind the horizon. 
%



 In \Fig{fig_Deq1np050Nmaxeq250_combine_plots_25Dec2024_TempBHvsEBH_26Dec2024:BHW}(right) we repeat the curves
 $S(\rho_{BH}), \bar{E}_{BH}, \bar{E}_{R}, T_{BH}$ (with the same colors on the left), 
 but now add the approximations to the temperature $T^{(approx)}_{BH}$ (the dotted magenta curve hugging the gray dashed $T_{BH}$ curve) 
 for early times $T^{(approx)}_{BH,<}$($N~\in~[0,177]$),
  and late times  $T^{(approx)}_{BH,>}$ ($N\in[177,251]$).
We find approximate fits given by

\bea{T:fits:BHW}
T^{(approx)}_{BH,<}(x) &\simeq& 1.35\,\left(\bar{E}_{BH}(0) - \bar{E}_{BH}(x)\right) 
\simeq 0.15\,x, \quad x\in[0,139], \label{T:fits:BHW:earlytimes} \\
T^{(approx)}_{BH,>}(x) &\simeq& 4\, \bar{E}^{1/8}_{BH}(x) \appropto e^{-0.215\,x}, \qquad \hspace{0.8in} x\in[178,251], \label{T:fits:BHW:latetimes} 
\eea
where again $x$ is the continuous, interpolated ``time" formed from the discrete evolution time $N$ in the One Shot mechanism.

The large spike in the BH temperature $T_{BH}$ (gray dashed curve) around $x_{Page}=163$ occurs just as the BH entropy $S(\rho_{BH})$ has zero slope, and hence begins to ``roll over." The presence of such a temperature spike at the Page time  is already anticipated from the generic Page curve \Fig{fig:PageCurve}, without assuming \tit{any} specific BH evaporation model, since from \Eq{Temp:defn:BHW} the denominator $dS_{BH}/dt$ goes to zero at $t_{Page}$ at point where the slope in the numerator $dE_{BH}/dt$ is non-zero (the slope $dE_{BH}/dt\to0$ only at the end stages where the BH has nearly completely evaporated, $t\to x\ge 200$ in \Fig{fig_Deq1np050Nmaxeq250_combine_plots_25Dec2024_TempBHvsEBH_26Dec2024:BHW}). After the Page time, there is a rapid (exponential) drop in both the entropy  $S_{BH}$  and mass/energy $\bar{E}_{BH}$ of the BH, as revealed in
\Fig{fig_Deq1np050Nmaxeq250_combine_plots_25Dec2024_TempBHvsEBH_26Dec2024:BHW}.

\subsubsection{$\mathbf{D=3}$ \tbf{and} $\mathbf{D=4:}$}\label{subsubsec:Deq1} 
\begin{figure}[h]
\begin{tabular}{ccc}
\hspace{-1.0in}
%
\includegraphics[width=4.25in,height=2.5in]{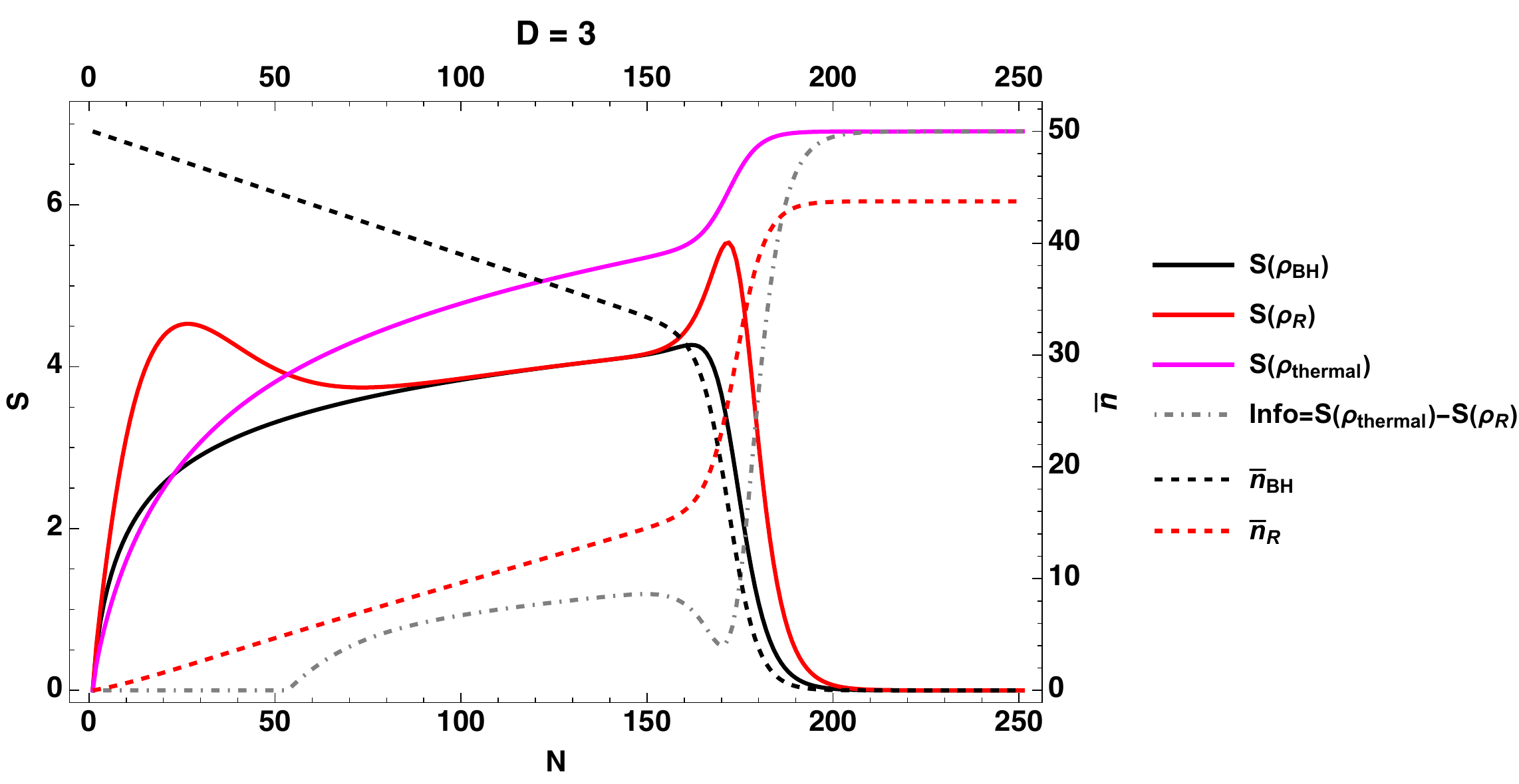} &
{}& 
\hspace{-0.35in}
\includegraphics[width=4.0in,height=2.4in]{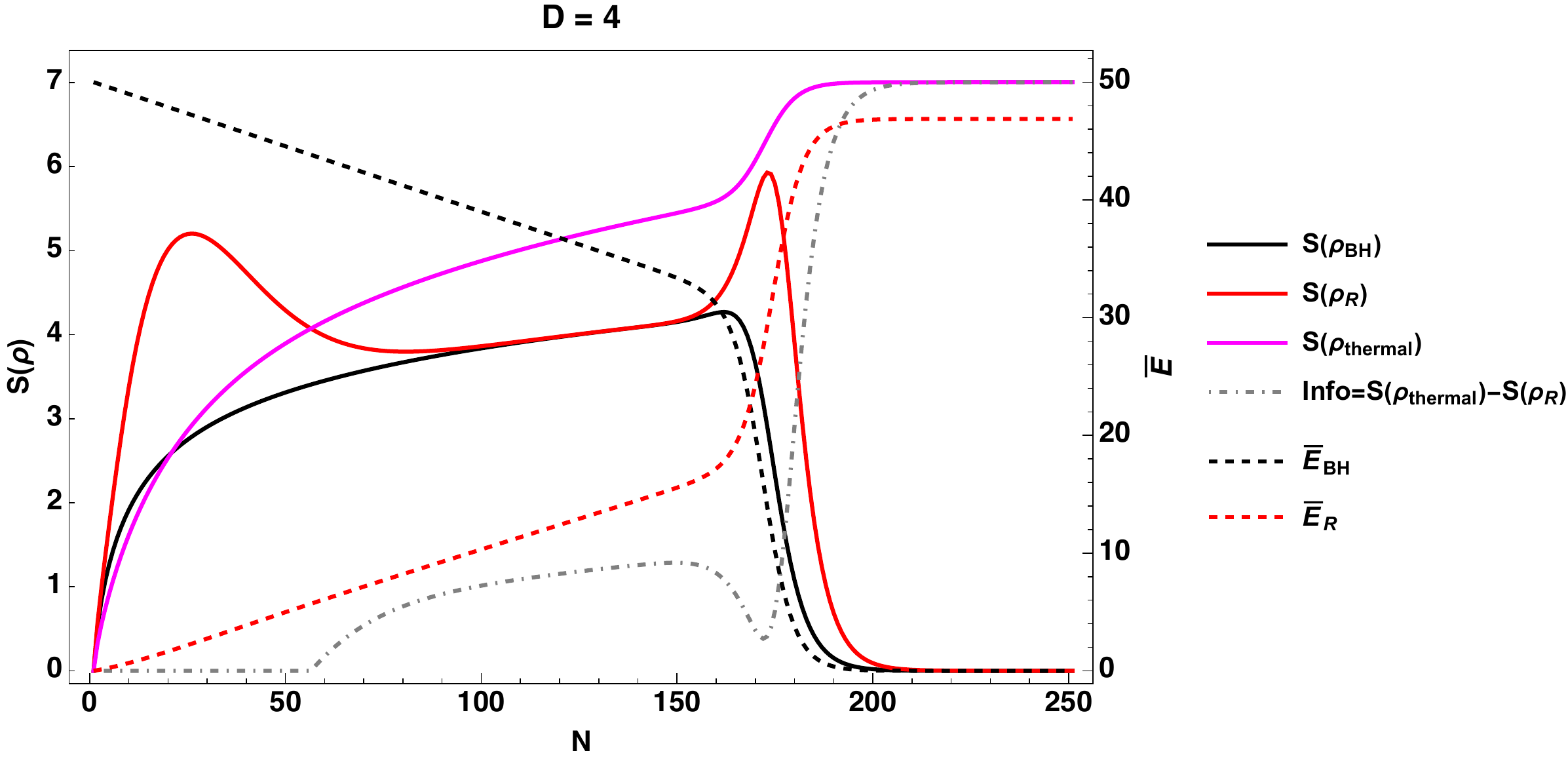}
%
\end{tabular}
\caption{Black hole entropy (black solid), Hawking radiation entropy (red solid),
effective Hawking radiation thermal entropy (magenta solid),  
(Hawking radiation) Page Information plots (gray dot-dashed) 
black hole energy (black dashed), Hawking radiation energy (red dashed)
for (left) $D=3$ and (right) $D=4$ with $\np0=50, z=0.1$.
Note that the energy of the Hawking radiation $\bar{E}_R$ asymptotes to 
$\left(1-\left(\thalf\right)^D\right)\,\np0\,\om_p = \{43.75,46.875\}$ for the left and right plots, respectively 
(setting $\om_p\to 1$), otherwise the graphs are qualitatively similar. 
The solid and dashed black lines are exactly the same as that in 
 \Fig{fig_Deq1np050Nmaxeq250_combine_plots_25Dec2024_TempBHvsEBH_26Dec2024:BHW} for $D=1$
(see \Eq{rhoBH:rhoR:BHW:rhoBH}).
}\label{fig_Deq3and4np050Nmaxeq250_combine_plots_15Dec2024:BHW}
\end{figure}

 In \Fig{fig_Deq3and4np050Nmaxeq250_combine_plots_15Dec2024:BHW} we plot the same curves as in 
 \Fig{fig_Deq1np050Nmaxeq250_combine_plots_25Dec2024_TempBHvsEBH_26Dec2024:BHW}(left, $D=1$), now for $D=3$ and $D=4$, again with $\np0=50, z=0.1$.
Note that the solid and dashed black lines are exactly the same as that in 
 \Fig{fig_Deq1np050Nmaxeq250_combine_plots_25Dec2024_TempBHvsEBH_26Dec2024:BHW} for $D=1$
 since $\rho_{BH}$ and hence $S_{BH}$ is independent of $D$
(see \Eq{rhoBH:rhoR:BHW:rhoBH}). One of the key differences in 
\Fig{fig_Deq3and4np050Nmaxeq250_combine_plots_15Dec2024:BHW} is that the
energy of the Hawking radiation $\bar{E}_R$ (red dashed curves) asymptotes to 
$\left(1-\left(\thalf\right)^D\right)\,\np0\,\om_p = \{43.75,46.875\}$ (setting $\om_p\to 1$)
for the left and right plots, respectively. In the limit $D\to\infty$ the entire BH initial mass/energy $\bar{E}_{BH}(0) = \np0\,\om_p $ would end up in the Hawking radiation  $\bar{E}_{R}$. Again, in the interior of the BH at the end of the evaporation process, there would remain $\np0$ idler/Hawking partner particles of total energy 
$\left(\thalf\right)^D\,\np0\,\om_p\to 0$ as $D\to\infty$. Thus, all the while there remains a finite number of particles in the interior of the BH as it evaporates that can be entangled with the exterior Hawking radiation, but end in the limit of vanishingly small mass/energy when the BH has evaporated. 

The other differences between the $D=1$ and $D=(3,4)$ is that the Page Information curve (gray dotdashed curves in \Fig{fig_Deq3and4np050Nmaxeq250_combine_plots_15Dec2024:BHW}) are 
(i) essentially zero longer for early times ($0\lesssim x\lesssim 50$), and 
(ii) rises less steeply for longer times ($50\lesssim x\lesssim 155$)  
than in the  $D=1$ case (gray solid curve 
in\Fig{fig_Deq1np050Nmaxeq250_combine_plots_25Dec2024_TempBHvsEBH_26Dec2024:BHW}).
In all $D$ cases, the Page Information rises exponentially around the crossover
$\bar{E}_{BH}= \bar{E}_{R}$ (intersection of the dashed black and dashed red curves), which is expected from the Page Curve. If we were to plot $\SBekHawk\equiv 2\,\pi\,\bar{E}^2_{BH}$  (not shown), the cross over in the energies occurs slightly before the crossover in $\SBekHawk = S_{BH}$ 
\footnote{For $D=3$, the energy crossover $\bar{E}_{BH}= \bar{E}_{R}$ and entropy crossover 
$\SBekHawk = S_{BH}$ occur at the approximately the same temporal point $x$ if $\SBekHawk$ were defined instead as  $\SBekHawk\approx (2\,\pi/3)\,\bar{E}^2_{BH}$}.
From the Page curve \Fig{fig:PageCurve}, both crossover conditions could be used to define the Page time $t_{Page}$.

An additional difference between 
\Fig{fig_entropies_Ebars_Dimeq3np0eq20tmaxeq1dteq0p1_15Dec2024:BHW} (the BHW model without the One Shot mechanism; direct numerical integration)
\Fig{fig_Deq1np050Nmaxeq250_combine_plots_25Dec2024_TempBHvsEBH_26Dec2024:BHW}, 
\Fig{fig_Deq3and4np050Nmaxeq250_combine_plots_15Dec2024:BHW} (the BHW model with the One Shot mechanism) is that in the former the initial slope of the BH energy is zero, $d\bar{E}_{BH}(0)/dt =0$,
while it falls linearly in the latter two plots. While both employ the waterfall mechanism, the latter employs the One Shot mechanism, while the former does not. We suspect that this most likely due to the approximations employed in the probabilities, 
 particularly in  the specific form, and number of denominators in the probabilities kept, 
  in \Eq{P:kd:kdm1:BHW} .
  (Note that in 
  \Fig{fig_Deq1np050Nmaxeq250_combine_plots_25Dec2024_TempBHvsEBH_26Dec2024:BHW}, and 
 \Fig{fig_Deq3and4np050Nmaxeq250_combine_plots_15Dec2024:BHW}  we kept all the denominators in 
  \Eq{P:kd:kdm1:BHW},  vs a maximum of $i_{max}=50$ used in \Fig{fig:Allcurves:n25:n100:BHW} with  
  $\np0=25$ and $\np0=100$). We return to this point in the Discussion \Sec{sec:Discussion}.
  
It is interesting, and not well understood the origin of the ``peaking" behavior in 
$S_R$ (solid red curve)  in both plots in 
\Fig{fig_Deq3and4np050Nmaxeq250_combine_plots_15Dec2024:BHW} occurring 
around $x\approx 25$ and $x\approx 173$. 
The former appears to occur when $S_{thermal} = S_{BH}$ (intersection of solid magenta and solid black curves), and the later seems to occur at an inflection point in $S_{thermal}$ (solid magenta curve) round the Page time. Also curious is that for a long stretch of time in between these peaks ($100\lesssim x\lesssim 150$), and just before the Page time, the entropy of the BH and Hawking radiation are almost equal $S_{BH}\approx S_{R}$. A more involved run for $D=5$ (not shown) shows essentially the same features as $D=(3,4)$, with these peaking behaviors slightly more pronounced. 
The behaviors of the energies, entropies, temperature and Page Information in 
 \Fig{fig_Deq1np050Nmaxeq250_combine_plots_25Dec2024_TempBHvsEBH_26Dec2024:BHW}  and
\Fig{fig_Deq3and4np050Nmaxeq250_combine_plots_15Dec2024:BHW} 
consist of one of the main results of this work.

\section{Summary and contribution of this present work}\label{sec:Summary}
The development of three models presented here are as follows.
In the first, trilinear Hamiltonian model \cite{Alsing:2015} the central feature was that the evaporating BH could be modeled as the depleting pump in fully quantized model of spontaneous parametric down conversion (SPDC). Thus, the entangled signal/idler pairs $\leftrightarrow$ external Hawking radiation/internal Hawking partner particles were created at the expense of the mass/energy of the pump/BH. 
From the evolution of a pure state in the Schr\"{o}dinger picture, this model led to a simple set of coupled quantum amplitude equations, which could be readily integrated. 
The advantage of this model, was that for a large initial BH population $\np0$ (assumed to be in a pure Fock state, only for simplicity of discussion and numerical computation) the state at early times was nearly a separable product of the pump and the emitted two-mode squeezed vacuum states. This is essentially the Hawking-like result for a constant mass BH (akin to a non-depleting pump source for SPDC). 
At later times, correlations and entanglement builds up between a depleting pump/BH and the interior/exterior Hawking pairs.
Key to this approach was that the initial number of particles $\np0$ modeling the BH was large but \tit{finite}, so that the BH population could be seen to deplete as the evaporation process proceeded.
However, because the Hamiltonian is Hermitian, after a time when the mean number of particles $\nbarp$ in the pump/BH is on the order of the emitted Hawking radiation $\nbars$, the latter begins to act like the pump, and the population in the pump/BH can build back up (like a laser in a cavity). Before this time though, the entropy curve of the BH and of the Hawking radiation does turn over (i.e. begins to decrease after its initial rise). However, the calculation should not really be taken literally after the point in time ($t>0$) at which $d\nbarp/dt=0$, after which the pump/BH energy could begin to rise due. This is due to the generated Hawking pairs (now with population/energy on par with the BH) acting back on the BH, now as the dominant driving SPDC ``pump source." 

To rectify the deficiencies of the first model while retaining its central features, the One Shot mechanism was introduced in \cite{Alsing:2016}. The purpose of this method was to mimic more realistically the escaping of the Hawking radiation to infinity so that the generated pairs could not act back on the pump/BH at longer times, thereby ensuring a monotonic loss in the BH energy. This was achieved by a Trotterization (temporal discretization) of the trilinear Hamiltonian so that it acted for a short time on a particular vacuum mode at a given time step, thus generating essentially squeezed state vacuum signal/idler pairs, before creating new new sets of signal/idler pairs from a different vacuum mode in the next time step. This mimicked the escaping of the signal/Hawking radiation to infinity.
The net effect of the One Shot method was that it did create entropy curves that 
(i) at early time when $\nbarp\gg\nbars$ had a thermal behavior (and essentially zero Page Information), 
while at later times 
(ii) created an effective Page time when $\nbarp\approx\nbars$ after which the entropy curves of the BH and Hawking radiation turned over and rapidly decreased to zero. During the entire evolution the system remained in a pure state, evolving under unitary evolution.
The model also demonstrated a spike in the temperature defined as $T\equiv dE_{BH}/dS_{BH}$ at this Page time turnover point. 

While having almost all the features desired of a BH evaporation model, the trilinear One Shot model had one glaring deficiency. Assuming (for simplicity of discussion and without loss of generality) degenerate SPDC where the each particle of the emitted Hawking pair has half the energy of the pump/BH, the end state of the system was $\nbarp=0$ while $\nbars=\nbari=\np0$. 
Thus, only half $\bar{E}_s =\thalf \np0\,\om_p$ the initial energy of the BH  $\bar{E}_p(0) = \np0\,\om_p$
 escapes to infinity in the Hawking radiation, 
 while the other half $\bar{E}_i =\thalf \np0\,\om_p$ remains behind the horizon in the idler/Hawking partner particles. This seems to imply an undesirable remnant as the end state of the BH evaporation process.
 
 To rectify the above  deficiency of the trilinear One Shot model, the ``waterfall" process was introduced as the main focus of this work. We desire the interior idler/Hawing particle partners to facilitate two processes, (i) promote the decrease in energy of the BH, and (ii) ``transport" energy from inside the BH to the exterior Hawking radiation. The central physical assumption introduced in the waterfall process is that each interior idler particle can act as its own pump source for SPDC. Thus,  an idler particle $i_1$ of energy $\om_p/2$ (created initially from the a pump/BH particle of energy $\om_p$) can further create ``downstream" signal/idler pairs 
 $(s_2,i_2)$ of half the energy $\om_p/2^2$ of the idler source $i_1$. In turn, the idler particle $i_2$ of energy $\om_p/2^2$ can create further signal/idler pairs  $(s_3,i_3)$ of energy $\om_p/2^3$, and the process continues ad infinitum. This process is also unitary (a sum of trilinear Hamiltonians). The pure state wavefunction admits a final state in which (i) the BH has zero energy, $(\nbarp(\infty)=0)\,\om_p=0$, while (ii) the Hawking radiation carries away $(1-(\thalf)^D)\,\om_p$, with $\np0$ signal particles in each mode 
 $(s_1,s_2,\ldots,s_d,\ldots,s_D)$, each of energy $(\thalf)^d\,\om_p$, and (iii) finally $0$ particles (energy) in each of idler modes $(i_1,i_2,\ldots,i_d,\ldots,i_{D-1})$, while $\np0$ particles in mode $i_D$ of energy 
 $(\thalf)^D\,\om_p$. In the limit $D\to\infty$, all the initial energy $\np0\,\om_p$ of the BH is carried away in the Hawking radiation, and the BH is completely evaporated, with no energy remaining in either the BH mode $p$ or all the interior idler/Hawking partner particle modes.
 The zero energy remaining in the interior Hawking partner particles in the 
 BH Waterfall model is qualitatively reminiscent of (and was inspired by)  the ``soft-hair" (i.e. near zero energy modes) enhanced entanglement qubit model of Hotta, Nambu and Yamaguchi \cite{Hotta_Nambu_Yamaguchi:2018}, but now employing a SPDC squeezing-based Hamiltonian.
 
 
 The BH One Shot Waterfall  model retains all the desirable features of the ($D=1$) One Shot model, 
while also creating Page Information curves that remain flatter (essentially zero ) for longer initial times, mimicking the desired thermal-like nature of the external signal/Hawking radiation. In both the One Shot and One Shot Waterfall model, the signal/idler states are composed of highly correlated (entangled) states themselves, which represents a superposition of  the vast number of ways that all the particles emitted by the pump(s) can be distributed into $N$ time slots at a given discrete One Shot evolution time $N$. The pure state of the composite system is then the pump/BH correlated with such emitted signal/idler, exterior/interior Hawking particle states. 

\section{Discussion and Future Work}\label{sec:Discussion}
In the Introduction we gave an overview of the modern view of BH evaporation and the BH Information Problem (Paradox). This view introduces many recently developed concepts such as  Holography, the Ryu-Takayangi (and HRT) formula, Quantum Extremal Surfaces, the Island Effect, Replica Wormholes, and Entanglement Wedge Reconstruction,  all of which have been gathering strong evidence over the last two decades  for computational  validity, and possible physical interpretation. While these developments have emerged out of approaches to realize the Page curve in BH evaporation, one drawback to the Euclidean path integral approach involved is that the method does not give the BH, nor the Hawking radiation, density matrix explicitly. 
Rather, it computes the von Neumann entropy directly $S(\rho) = -\Tr[\rho\log\rho]$ (via the Euclidean Path Integral Replication Method).

One advantage of the BH One Shot Waterfall model described in this work is that it does provide a pure state wavefunction (approximation) for the composite BH, (all) signals/idlers $BH\cup S \cup I$ system, from which the von Neumann entropy (and other entropies and relevant quantities) can be computed explicitly. 
It might be argued that its one major drawback is that it is totally devoid of geometric content (no semiclassical gravitational background is assumed, nor apparent in the its derivation).
However, we would argue that the geometry is, in sense, implicit in this model, 
through the use of the SPDC process. 
As discussed in the introduction, SPDC and the generation of two-mode squeezed vacuum states (i.e. signal/idler Hawking pairs), appears naturally for accelerated observers in both flat spacetime (Unruh effect), as well as stationary spacetimes (Hawking effect), both with a temporal Killing vector. The squeezing occurs due to the  accelerated collapsing surface of the BH horizon, akin to an accelerated mirror \cite{Birrell_Davies:1982}, and hence is generic in a BH evaporation process, one could argue, essentially independent of the geometry (recall the generation of a squeezed state from  a simple harmonic oscillator rapidly changing its frequency under the sudden approximation, discussed on page 10).
Another advantage of models based on a trilinear Hamiltonian, is that they hold the prospect of being experimentally testable, as evidenced by the recent thesis of Guti\'{e}rrez
 \cite{Gutierrez_phdthesis:2022}, which was directly inspired by the work of Br\'{a}dler and Adami \cite{Bradler_Adami:2016}.
The goals of the BH One Shot Waterfall model were to capture the essential features of the BH evaporation process, and be able to naturally reproduce a Page curve for the resulting entropies. This model appears to accomplish these goals.

As to any direct connection of the BH One Shot Waterfall model to the modern viewpoint of BH evaporation, one can only speculate at this point. 
One of the central questions of black hole evaporation is the question of where exactly do the degrees of freedom reside that make up $\SBekHawk$. One possible view (see the review by D.N. Page \cite{Page:2005}) is that in a full theory of quantum gravity one might not expect a definite four-metric, or even a definite causal structure. This could in turn blur the absolute distinction between the inside and outside of the BH. There still remains the difficult question of how the information of those degrees of freedom can get out as the BH evaporates.
It is curious that the function of the Replica Wormholes and the formation of the interior Island in the modern view is to associate a portion of the interior Hawking partner particles with the external Hawking radiation, so that as time evolves, much of the interior particles are moved to the exterior until the entropy of the BH decreases as the Bekenstein-Hawking formula $\SBekHawk = \tfourth Area(BH)$.
In the BH One Shot Waterfall model a seemingly related functionality is achieved by the idlers/Hawking partner particles acting as subsequent SPDC pump sources. The latter effect is negligible at early times, and only contributes once substantial intermediate population build up in the idlers (all the while they are being simultaneously bled away  downstream by their own creation of signal/idler pairs). This appears qualitatively  ``reminiscent" of the buildup of the Island effect in the modern view, but a direct quantitative comparison is yet to be made (if even possible).

While a full theory of quantum gravity remains as of yet elusive,
the robust features of the semiclassical BH evaporation calculation by Hawking (and Unruh)  should emerge in the low energy limit. Since the process of squeezing is at the heart of these semiclassical calculations, the bilinear Hamiltonian of the optical SPDC process serves as reasonable phenomenological approximation when the BH mass is either fixed, or considered so large as essentially to be constant \cite{Adami_VerSteeg:2014}. The next level of approximation to attempt to incorporate some level of back-reaction in the BH evaporation process, would be to allow the mass of the BH to decrease. The trilinear model involved here, and studied by previous authors \cite{Alsing:2015,Alsing:2016,Bradler_Adami:2016,Nation_Blencowe:2010,Gutierrez_phdthesis:2022} is the simplest, and most natural Hamiltonian extension of the prior bilinear SPDC model. One could reasonably argue that a low-energy approximation of any effective Hamiltonian arising from a consistent theory of quantum gravity must incorporate some form of a tripartite interaction between the BH, the emitted Hawking radiation and their 
infalling partner particles, and hence would be most simply phenomenologically modeled by a trilinear Hamiltonian of SPDC with a depleting ``pump." Such low energy calculations would be kind of calculations that would first be performed if a consistent theory of quantum gravity existed.

For future work, several further avenues of the BH One Shot Waterfall model can be explored. In this work we took the rapidity parameter as $z=0.1$. One could also work with a temporal profile for $z=z(N)$, as was explored in the $D=1$ One Shot model in \cite{Alsing:2016} in order to assess its modification to the Page curves (e.g. attempting to make the Page Information curve nearly zero for longer periods of time, even up to just before the Page time). The model presented here also does not follow exactly the historical (Hawking) power law rate of decay for the mass of the BH \cite{Hartle:2003, Carroll:2004}, $dM(t)/dt\simeq -1/M(t)^2$ (where here, $\bar{E}_{BH} = M$ for $c=1$). Instead the mass/energy falls approximately linearly for early time before the Page time, and exponentially after the Page time. This issue could be explored further (through possible modifications to the temporal profile of $z$ in an attempt to create such a power law decay). Further, a deeper understanding of the spiking effects in the entropy of the Hawking radiation in \Fig{fig_Deq3and4np050Nmaxeq250_combine_plots_15Dec2024:BHW}, and its near equality with the BH entropy for long stretch of time before Page time (and its implication on the nature of the wavefunction during this time period) is warranted. (Interestingly, in \cite{Alsing:2015} it was noted that the state of the BH was essentially a single mode squeezed state (witnessed by the probability for odd occupation numbers to be close to zero), at the energy cross over point, where the signal/idlers start significantly acting back on the pump/BH). Additionally, faster computational power could check the behavior of the entropy and associated relative curves for much larger values of $D$ (the computationally more time consuming $D=5$, not shown, was explored, but revealed no significant changes from the $D=(3,4)$ cases).
Lastly, using a pure Fock (number) state for the initial state of the pump was made purely for descriptive and computational reasons, without loss of generality.  It is straightforward to create an arbitrary initial state for the BH by simply include an additional sum over $\np0$, with associated initial state probabilities. A natural choice would be to put the BH in a coherent state (as was done in  \cite{Alsing:2015}), as this represents the quantum state most like a classical laser (and hence, a classical BH). But other random states could also be easily incorporated. Since the model also lends itself to initially seeding the initial signal, so as to represent in-falling matter (as in  \cite{Alsing:2015}), this could be used to explore phenomena such as the Hayden-Preskill BH scrambling time \cite{Hayden_Preskill:2007}, which has relevance to the time at which the island first begins to appear 
(see \cite{Almheiri_Hartman:2020, Almheiri_Hartman:2021, Almheiri:2020, Penington:2022}).
These and other avenues will be explored in future work.

\ack The author would like to acknowledge N. Natek for useful discussions.
The author has  no competing interests for this work.
The \tit{Mathematica} codes that support the data used in this paper and other finding are available from the  author upon reasonable request and acknowledgement of use.
\appendix
\section{Derivation of One Shot Black Hole Waterfall model}\label{app:BHW:derivation}
In this Appendix we derive the approximate One Shot Black Hole Waterfall state 
$ \ket{\psi^{(N)}_D} $ in \Eq{psi:N:D:BHW}, 
with probabilities $P_{k_{d}}^{(k_{d-1})}$ given in \Eq{P:kd:kdm1:BHW}.

As in the One Shot model described in \Sec{subsec:OneShot:2016}, we allow the pump/BH to evolve for a short amount of time to create signal/idler pairs out of the vacuum, before ``moving on" (in the next step of the Trotterization) to create new pairs from new vacuum modes (as the BH horizon shrinks in radius as it evaporates). This involves the bare probabilities such as
\Eq{Psi:1:exact:probs} and \Eq{Psi:2:approx:BHW}. The BH Waterfall model essentially entails adding a depth $D$ cascade of idler pump sources at each time step of the depth $D=1$  original One Shot model. For this we need to introduce a double-index set of integers $\{n_{i,d}\}$ denoting the number of emitted particles at the One Shot evolution time step $i\in\{1,\ldots,N\}$ (indexing the BH emission events), at  intermediate depth $d\in\{1\,\ldots, D\}$ (indexing the subsequent idler $i_d$ emission events). The generalization $\ket{\Psi^{(N)}_D}$  of $\ket{\Psi(N)}$ in \Eq{Psi:N:BHW} then becomes
\bea{PsiND:p8.3:BHW}
\fl
&{}& \hspace{-0.25in}\ket{\Psi^{(N)}_D}=  \no
%
\fl
& & \sum_{n_{11}=0}^{\np0} \sum_{n_{12}=0}^{n_{11}} \sum_{n_{13}=0}^{n_{12}} \cdots \sum_{n_{1D}=0}^{n_{1(D-1)}}
\left[
P_{n_{11}}^{(\np0)} P_{n_{12}}^{(n_{11})} P_{n_{13}}^{(n_{12})}\cdots P_{n_{1D}}^{(n_{1(D-1)})}\,
\right]^{1/2}\, \no
\fl & & \hspace{1.0in}
\ket{n_{11}-n_{12}}_{i_{11}} \ket{n_{12}-n_{13}}_{i_{12}}\cdots\ket{n_{1D}}_{i_{1D}} 
\ket{n_{1D}}_{s_{1D}} \ket{n_{1(D-1)}}_{s_{1(D-1)}}\cdots\ket{n_{11}}_{s_{11}} \otimes \no
%
\fl
& & 
 \sum_{n_{21}=0}^{\np0-n_{11}} \sum_{n_{22}=0}^{n_{21}}  \sum_{n_{23}=0}^{n_{22}}\cdots \sum_{n_{2D}=0}^{n_{2(D-1)}}
 \left[
P_{n_{21}}^{(\np0-n_{11})} P_{n_{22}}^{(n_{21})} P_{n_{23}}^{(n_{22})} \cdots P_{n_{2D}}^{(n_{2(D-1)})}\,
\right]^{1/2}\, \no
%
%
\fl & & \hspace{1.0in}
\ket{n_{21}-n_{22}}_{i_{21}} \ket{n_{22}-n_{23}}_{i_{22}}\cdots\ket{n_{2D}}_{i_{2D}} 
\ket{n_{2D}}_{s_{2D}} \ket{n_{2(D-1)}}_{s_{2(D-1)}}\cdots\ket{n_{21}}_{s_{21}} \otimes \no
%
%
\fl & & \vdots \no
%
%
\fl
& & 
 \sum_{n_{N1}=0}^{\np0-(n_{11}+n_{21}+\cdots+n_{N1})} \sum_{n_{N2}=0}^{n_{N1}} \sum_{n_{N3}=0}^{n_{N2}}\cdots \sum_{n_{ND}=0}^{n_{N(D-1)}}
\left[
P_{n_{N1}}^{(\np0-(n_{11}+n_{21}+\cdots+n_{N1}))} P_{n_{N2}}^{(n_{N1})}\cdots P_{n_{ND}}^{(n_{N(D-1)})}\,
\right]^{1/2}\,
\no
\fl & & \hspace{1.0in}
\ket{n_{N1}-n_{N2}}_{i_{N1}} \ket{n_{N2}-n_{N3}}_{i_{N1}}\cdots\ket{n_{ND}}_{i_{ND}} 
\ket{n_{ND}}_{s_{ND}} \ket{n_{N(D-1)}}_{s_{N(D-1)}}\cdots\ket{n_{N1}}_{s_{N1}} \otimes \no
%
%
\fl & & \hspace{1.0in} \ket{\np0 - (n_{11}+n_{21}+n_{31}+\cdots+n_{N1})}_{p},
\eea
where the probabilities are given by
\be{p8.3:probs:BHW}
P_{n_{emit}}^{(n_{initial})} = \frac{1-z}{1-z^{n_{initial}+1}}\,z^{n_{emit}}, \qquad 
\sum_{n_{emit}=0}^{n_{initial}} P_{n_{emit}}^{(n_{initial})} =1,
\ee
where $n_{initial}$ is the initial number of particles in the idler pump, and $n_{emit}$ is the 
number of particles it emits into the next signal/idler pair (with each emitted particle having half the initial energy of the idler pump that created it).
We can think of the above summations and probabilities as a two-dimensional grid labeled by $i\in[1:N]$ (running down the page) as the rows and $d\in[1,D]$ (running across the page) as the columns. The important point is that each probability has a power of $z^{n_{i d}}$, so that the total factor of $z$ is given by
\bea{zs:BWH}
z^{\sum_{i=1}^{N} \sum_{d=1}^{D} \, n_{i,d}} &=& 
z^{\sum_{i=1}^{N} n_{i 1}}\,z^{\sum_{i=1}^{N} n_{i 2}} 
                                         \,z^{\sum_{i=1}^{N} n_{i 3}}\cdots z^{\sum_{i=1}^{N} n_{i,D}}, \no
%
&\equiv& z^{k_1}\, z^{k_2}\,z^{k_3}\,\cdots  z^{k_D},
\eea
where we have ``summed down" each column labeled by $d$, and introduced the collective total-emission variables $k_d \equiv \sum_{i=1}^{N} n_{i d}$, representing the total number of particles emitted by idler pump $i_{d-1}$ in $N$ times steps (where $i_0\equiv p$ is the BH pump). Approximating all the denominators 
in each $P_{n_{emit}}^{(n_{initial})}$ in \Eq{p8.3:probs:BHW} by unity gives the collective unnormalized states signal/idler states $\ket{\tilde{\Phi}^{(N)}_{k_1,k_2,\ldots,k_D}}$. 
These states contain the rest of the summations over the remaining collective variables defined as 
$j_{i,d}~\equiv~\sum_{i'=0}^{i} n_{i',d}$, generalizing the sums 
int the state $\ket{\tilde{\Phi}^{(N)}_{k_1}}$
in \Eq{Psi:j:jN:first:BHW} for the depth $D=1$ case.
As in the original $D=1$ One Shot method, we can create a finer 
approximation to these denominators by using their lowest power of $z$ (since we assume $z\ll 1$), and approximate them by their largest contributions. This yields the factor of $\frac{1}{(1-z^{k_{d-1}-k_{d}+1})^{N-d}}$ in \Eq{P:kd:kdm1:BHW}. 
Essentially, the factor $\frac{1}{(1-z^{\np0-k_{1}+1})^{N-d}}$ at depth $d=1=D$ goes to
 $\frac{1}{(1-z^{k_{d-1}-k_{d}+1})^{N-d}}$ at intermediate depth $d\in[1,D]$, with one caveat.
 The former factor can be pulled all the way to the left (in front of $\sum_{k_1}^{\np0}$) in the nested summations $\sum_{k_1}^{\np0}\sum_{k_1}^{k_2}\sum_{k_2}^{k_3}\cdots\sum_{k_D}^{k_{D-1}}$
 over the collective emission variables $(k_1,k_2,k_3,\ldots,k_D)$, since it only depends on $k_1$. 
 However, subsequent factors $\frac{1}{(1-z^{k_{d-1}-k_{d}+1})^{N-d}}$ 
 for the idler pump at intermediate depth $d$ 
 can only be pulled as far to left as the summation $\sum_{k_d}^{k_{d-1}}$ (hence the power $N-d$ now instead of $N-1$) 
 \footnote{Formally we should use the power $N-d$ if it is non-negative, and use, say $N-1$ if $N-d<0$, since the (discrete) time steps are $N\ge 1$.
However, since $z<<1$ we have found that numerically there is not much effect in the early time steps if we just continue to use the power $N-d$. A conditional ``if-else" coding statement, 
such as $\textrm{If}[N-d\ge 0, N-d, N-1]$, tends to slow down the numerical computations.}.
 
 Normalizing the state  by the inverse square root of the number of states created at each depth $d$ yields the binomial factors 
 $
 {\tiny
 \left(
       \begin{array}{c}
          k_d + N-1 \\
          k_d
        \end{array}
\right)
}
$ 
in the probabilities
in \Eq{P:kd:kdm1:BHW}. This yields the final approximated state $\ket{\psi^{(N)}_D}$ in \Eq{psi:N:D:BHW} with the probabilities $P_{k_{d}}^{(k_{d-1})} $ for $d\in[1,D]$ in \Eq{P:kd:kdm1:BHW}.
Simply viewed, \Eq{P:kd:kdm1:BHW} for a general ``waterfall" depth $D$, is just a nested iterated version of the single depth $D=1$ One Shot model in \Eq{Pksi:corrected:BHW:1}, with the probabilities 
in \Eq{Pksi:corrected:BHW:2} generalized to  \Eq{P:kd:kdm1:BHW}, with one factor of the square root of the probability for each intermediate depth (idler pump) $d\in[1,D]$.
%
\section*{References}
\providecommand{\noopsort}[1]{}\providecommand{\singleletter}[1]{#1}%
\providecommand{\newblock}{}

\end{document}